\documentclass[12pt,preprint,epsf]{aastex}
\usepackage{natbib}

\slugcomment{ }

\begin{document}
\pagestyle{plain}
\pagenumbering{arabic}
\parindent = 0 mm
\received{31 October 2016}
\title{Optical Sky Brightness and Transparency during the Winter Season at Dome A Antarctica from the 
Gattini-Allsky Camera}
\author{
Yi Yang,\altaffilmark{1,2}
Anna M. Moore,\altaffilmark{3}
Kevin Krisciunas,\altaffilmark{1}
Lifan Wang,\altaffilmark{1,4,7}
Michael C. B. Ashley,\altaffilmark{5}  
Jianning Fu,\altaffilmark{2}
Peter J. Brown,\altaffilmark{1}
Xiangqun Cui,\altaffilmark{4,6}
Long-Long Feng,\altaffilmark{4,7}
Xuefei Gong,\altaffilmark{4,6}
Zhongwen Hu,\altaffilmark{4,6}
Jon S. Lawrence,\altaffilmark{8}
Daniel Luong-Van,\altaffilmark{5}
Reed L. Riddle,\altaffilmark{11}
Zhaohui Shang,\altaffilmark{4,9}
Geoff Sims,\altaffilmark{5}    
John W. V. Storey,\altaffilmark{5} 
Nicholas B. Suntzeff,\altaffilmark{1}
Nick Tothill,\altaffilmark{10}
Tony Travouillon,\altaffilmark{11}
Huigen Yang,\altaffilmark{4,12}
Ji Yang,\altaffilmark{4,7}
Xu Zhou,\altaffilmark{4,13} 
and Zhenxi Zhu\altaffilmark{4,7}
}
\altaffiltext{1}{George P. and Cynthia Woods Mitchell Institute for Fundamental
Physics \& Astronomy, Texas A. \& M. University, Department of Physics \& 
Astronomy, 4242 TAMU, College Station, TX 77843; {ngc4594@physics.tamu.edu}}

\altaffiltext{2}{Department of Astronomy, Beijing Normal University, Beijing 100875,
P. R. China}

\altaffiltext{3}{Research School of Astronomy and Astrophysics,
The Australian National University, Canberra, ACT 2611, Australia }

\altaffiltext{4}{Chinese Center for Antarctic Astronomy, Purple Mountain Observatory,
Chinese Academy of Sciences, Nanjing 210008, P. R. China}

\altaffiltext{5}{School of Physics, University of New South Wales, Sydney, NSW 2052,
Australia.}

\altaffiltext{6}{Nanjing Institute of Astronomical Optics \& Technology ,National
Astronomical Observatories, Chinese Academy of Sciences, Nanjing 210042,
P. R. China.}

\altaffiltext{7}{Purple Mountain Observatory, Chinese Academy of Sciences, Nanjing
210008, P. R. China.}

%
\altaffiltext{8}{Australian Astronomical Observatory, Sydney, NSW 2113, Australia.}

\altaffiltext{9}{Tianjin Astrophysics Center, Tianjin Normal University, Tianjin
300387, P. R. China.}

\altaffiltext{10}{School of Computing, Engineering \& Mathematics, University of
Western Sydney, NSW 2751, Australia }

\altaffiltext{11}{California Institute of Technology, Pasadena, CA 91125, USA.}

\altaffiltext{12}{Polar Research Institute of China, Shanghai 200136, P. R. China.}

\altaffiltext{13}{National Astronomical Observatories, Chinese Academy of Science,
Beijing 100012, P. R. China.}

\begin{abstract}
The summit of the Antarctic plateau, Dome A, is proving to be 
an excellent site for optical, NIR, and THz astronomical 
observations. GATTINI was a wide-field camera installed on the 
PLATO instrument module as part of the Chinese-led traverse to 
Dome A in January, 2009. 
{We present here the measurements of sky 
brightness with the Gattini ultra-large field of 
view ($90^{\circ} \times 90^{\circ}$)
in the photometric $B$-, $V$-, and $R$-bands, cloud 
cover statistics measured during the 2009 winter season, and 
an estimate of the sky transparency.} A cumulative probability 
distribution indicates that the darkest 10\% of the nights at 
Dome A have sky brightness of S$_B$ = 22.98, S$_V$ = 21.86, and 
S$_R$ = 21.68 mag arcsec$^{-2}$. These values were obtained 
around the {year 2009} with minimum aurora, and they are comparable 
to the faintest sky brightness at Mauna Kea and the best sites 
of northern Chile. Since every filter includes strong auroral 
lines that effectively contaminate the sky brightness 
measurements, for instruments working around the auroral lines, 
either with custom filters or with high spectral resolution 
instruments, these values could be easily obtained on a more 
routine basis. In addition, we present example light curves for 
bright targets to emphasize the unprecedented observational 
window function available from this ground-based site. 
These light curves will be published in a future paper.
\end{abstract}
\vspace{-2em}
\section{\vspace{-1em}INTRODUCTION}
Over the past centuries, people's growing demand for large 
astronomical facilities keeps pushing forward the progress 
of astronomical site selections. Some of the most major 
considerations for ground-based optical and IR astronomy 
include the seeing, atmospheric transparency and cloud 
coverage, number of clear nights, wind speed, precipitable 
water vapor, {and thermal backgrounds.} 
{
Various site surveys in recent years have revealed the advantages of the 
Antarctic plateau sites. Low and stable count rates of sky 
backgrounds in optical bandpasses have been measured at Dome C \citep{Kenyon_etal_2006b}, 
South pole \citep{Nguyen_etal_1996, Ashley_etal_1996}, and Dome A \citep{Zou_etal_2010, Sims_etal_2012b}. 
High atmospheric transmission has been inferred at Dome A \citep{Lawrence_2004, Yang_etal_2009} and various 
Antarctic sites \citep{Lawrence_2004}. Good average optical seeing above the boundary layer can be 
obtained at South Pole (i.e., 0$\arcsec$.37 in 17$-$27 m above the ground, \citealp{Marks_etal_1996}) 
and Dome C (i.e., 0$\arcsec$.27, \citealp{Lawrence_etal_2004}; 0$\arcsec$.36$\pm$0$\arcsec$.19 \citealp{Agabi_etal_2006}; 
0$\arcsec$.36, \citealp{Aristidi_etal_2009}; $\textless$0$\arcsec$.3, \citealp{Giordano_etal_2012}, 
in $\sim$30 m above the ground.; also see \citealp{Vernin_etal_2009, Aristidi_etal_2015}). 
So far the best seeing among all the ground-based astronomical sites 
was measured to be 0$\arcsec$.07 at Dome C \citep{Lawrence_etal_2004}.
Many astronomical observations could benefit from the 
consecutive periods of dark time at polar latitudes.} 
Comparisons between different astronomical sites among the 
Antarctic plateau being tested for the cloud coverage, 
aurorae, thickness of the boundary layer, seeing, humidity, 
and temperature {are} given by \citet{Saunders_etal_2009}. 

The high-altitude Antarctic sites of Dome A (latitude 
$80^{\circ} 22\arcmin$ S, longitude $77^{\circ} 21\arcmin$ 
E, elevation 4093m) offer intriguing locations for future 
large-scale astronomical observatories. Site testing {work} 
since 2008 {has} confirmed Dome A to be an excellent 
astronomical site. The extremely thin turbulent boundary 
layer measured to be 13.9 m near the ground at Dome A enables 
a free-of-atmosphere observing condition for a telescope on a 
small tower \citep{Bonner_etal_2010}. Some other advantages 
including the low sky brightness measured in SDSS $i$ band 
\citep{Zou_etal_2010}, the outstanding low cloud coverage 
compared to other astronomical sites \citep{Zou_etal_2010}, 
and the extremely low atmospheric water vapor content 
\citep{Sims_etal_2012a}. Additionally, the airglow and aurorae 
at Dome A within optical and near-IR range during the 2009 
winter season have been characterized by 
\citet{Sims_etal_2012a}, and only 2\% of the time during 2008 
winter season (Solar minimum) have shown strong aurorae events 
in i-band \citep{Zou_etal_2010}. Dome A also has exceptional 
transmission and multi-day persistent superlative observing 
conditions in {the} Terahertz regime \citep{Yang_etal_2010}. 

Time-series photometry has long been an essential tool to study 
the stellar properties as well as other astrophysical phenomena 
involving time-variant celestial objects. Long-term monitoring 
of stars to a very high degree of precision probes a wide range 
of frequencies. Over the last decades, there has been 
rapid progress in consecutive, high-quality, and high-cadence 
transiting surveys. Space-based missions including the Swift 
Ultraviolet/Optical Telescope \citep{Gehrels_etal_2004, 
Roming_etal_2005} is leading the high-energy regime, the CoRoT 
satellite \citep{Baglin_etal_2006} 
and the Kepler satellite \citep{Borucki_etal_2010} are searching 
for Earth-size planets and performing asteroseismology on the 
field stars. A summary of ground-based untargeted transient and 
variable surveys can be found at the Table~1 in 
\citet{Rau_etal_2009}. 

Consecutive monitoring for as long as months is not possible in 
single mid-latitude astronomical observatories. 
While space-based transient surveys achieve 
better accuracy due to the extremely low atmospheric absorption, 
turbulence, and light pollution, ground-based sites offer 
advantages such as unlimited cadence, flexible pointing, and the 
feasibility of following brighter targets. Taking the advantages 
of the long ``winter night'' as well as other remarkable 
observation conditions at Dome A, time-series carried out by small 
telescopes have already obtained data with high quality, opening 
a window for asteroseismology at Antarctica plateau sites. 

Previous works demonstrated that {high} photometric accuracy can 
be acquired by small aperture telescopes at Dome A. The first and 
comprehensive studies on asteroseismology and stellar physics at 
Dome A Antarctica have been conducted by the 14.5 cm diameter 
Chinese Small Telescope ARray (CSTAR, \citealp{Yuan_etal_2008}), 
which monitoring an area around the South Celestial Pole since 
2008. The nature of the CSTAR wide field design (FOV of 
$4.5^{\circ} \times 4.5^{\circ}$ and the absence of sidereal 
tracking system introduced significant systematic uncertainties 
into stellar photometry. {Various efforts have been made to reduce 
the systematic errors and to push the photometric precision below 
a few mmag, including the modeling of the inhomogeneous effects of 
clouds \citep{WangS_etal_2012}, the ghost images \citep{Meng_etal_2013}, 
and the systematic diurnal residuals \citep{WangS_etal_2014_RAA}. 
Based on the detrended light curves obtained during the 2008 winter 
season, comprehensive studies on exoplanet 
candidates \citep{WangS_etal_2014_ApJS}, stellar 
variability \citep{WangS_etal_2015}, eclipsing 
binaries \citep{Yang_etal_2015}, and stellar 
flares \citep{Liang_etal_2016} have been carried out}. 
Other independent studies include the variable 
sources \citet{Wang_etal_2011, Wang_etal_2013, Oelkers_etal_2015}, 
and specific studies on the pulsation modes of RR Lyrae 
\citep{Huang_etal_2015} and $\delta$ Scuti 
variables \citep{Zong_etal_2015} based on the CSTAR observations 
in single or multiple years from 2008, 2009, and 2010; 
the variable stars observed during 2012 winter 
season \citep{Li_etal_2015} with a single unit of the Antarctic 
Survey Telescopes (AST3, \citealp{Yuan_etal_2014}). 
In addition, time-series photometry studies have also been conducted 
at Dome C Antarctica, {for instance, careful time-series multi-color 
photometry to study the stellar pulsation and evolution using the 
Photometer AntarctIca eXtinction (PAIX, \citealp{Chadid_etal_2010, 
Chadid_etal_2014, Chadid_etal_2016}).} 

These transient surveys at Antarctic sites with $\sim$10 cm class 
and larger telescopes have measured the stars within a magnitude range 
of $\sim$8 to $\sim$15 with a FOV of $\lesssim20$ square degrees. 
The multi-band ultra-wide FOV imaging 
obtained by Gattini-Dome A camera during the 2009 winter season 
is also a valuable dataset considering its continuous monitoring 
of the flux variation of the 3-7 magnitude stars. 
Multi-band photometric results for bright targets 
obtained during {the} 2009 winter season by {the} 
Gattini-Dome A camera will be discussed in a future paper.

In this paper, we analyze a multi-wavelength dataset collected 
at Dome A, Antarctica, during the 2009 winter season. We have 
measured and calibrated the sky brightness in three photometric 
bands. We estimate the transparency variations and perform 
aperture photometry for those targets in the $V$ magnitude range 
$\sim$3.5 to $\sim$7.5. We focus on targets {in} the sky south 
of declination $-50^{\circ}$. 
The organization of this paper is as follows: In Section 2 we 
describe the instrument and observations. Section 3 discusses 
data reduction pipeline. In Section 4 we give our results, 
and in Section 5 our conclusions.
\vspace{-1em}
\section{\vspace{-1em}Importance of Sky Background Measurements}
When doing astronomical photometry, the ideal output is to 
determine the brightness of individual celestial objects. 
However, various sources including the {scattered} light 
from the Sun, the Moon, the aurora and airglow, will 
contaminate the flux from the astronomical sources. A 
summary of the contributions to the light of the night sky 
has been provided by \citet{Roach_etal_1973} and 
\citet{Kenyon_etal_2006b}. Considering the case of aperture 
photometry, the total flux integrated over the circular 
aperture can be expressed as $Flux = Source + Sky + RN^2 
+ Dark$. The terms on the right-hand side are the source 
counts enclosed by the aperture from a celestial object, the 
sky background, the readout noise ($RN^2$) , and the electron 
counts from the dark current, respectively. Due to the 
discrete nature of the electric charge, a Poisson process models 
the counting process of the photons, and the error is given 
by the square root of the total counts. By adding the noise 
terms in quadrature, the signal-to-noise (S/N) for object 
measured in aperture with a radius $r$ can be written as: 
$S/N = Source/ \sqrt{Source+Sky+RN^2+Dark}$. In a sky limited 
case, i.e., $\sqrt{Sky} \textgreater 3RN$, $S/N \approx 
Source / \sqrt{Sky}$. The temporal variation of the sky 
background significantly influences the efficiency and 
detection capability of ground-based astronomical facilities, 
especially in the low S/N regime. 

Given the critical role 
played by sky brightness in astronomical site selection, the 
measurement of the background light of the sky has been carried 
out for over a century using photographic plates, 
photomultiplier tubes, and modern digital detectors. A classic 
monograph on the subject is {\em The Light of the Night Sky} by 
\citet{Roach_etal_1973}. A comprehensive summary of the 
published sky background statistics is given by 
\citet{Benn_etal_1998}. Some useful background can also be found 
in papers by \citet{Walker_1988}, \citet{Krisciunas_1997}, 
\citet{Patat_2003}, \citet{Krisciunas_etal_2007_sky}, and references 
therein.  

The daily and monthly variations of the sky brightness at the 
high-altitude Antarctic Dome A site are expected to be 
different than mid-latitude sites. 
{
It can be shown that the angle of solar elevation, $\alpha$, 
can be approximated in terms of the Solar declination angle at 
a given date, $\delta(d)$, the hour angle of observation at 
a given time during the day, $h(t)$, and the observer's latitude, 
$\phi$: 
\begin{equation}
\alpha = \mathrm{sin^{-1} [sin } \delta(d) \mathrm{sin} \phi + \mathrm{cos} \delta(d) \mathrm{cos} h(t) \mathrm{cos}\phi]
\label{eqn_sunalt1}
\end{equation}
The declination angle can be calculated by: 
\begin{equation}
\delta = 23^{\circ}.45 \times \mathrm{sin} \bigg{[} \frac{2\pi}{365} \times (d+284) \bigg{]}
\end{equation}
Where $d$ is the day of the year with January 1$^{st}$ as $d=1$; given 
$\phi = -80^{\circ} 22 \arcmin$ of Dome A site, on the summer solstice ($d=172$), 
the Sun obtains its lowest mean altitude over a sidereal day, i.e., 
$h(t)$ from $0^{hr}$ to $24^{hr}$, $\alpha$ from $-13^{\circ}.8$ to $-33^{\circ}.1$. 
Therefore, even continual darkness can be 
expected during the winter season at Antarctic sites, 
however, the latitude of Dome A never allows the Sun to 
stay more than 18$^{\circ}$ below the horizon for an entire 
sidereal day.} 
The twilight due to the scattered light from the 
Sun and the Moon without the additional scattering by clouds 
need to be modeled to better understand the scattering effect 
of the atmosphere at Dome A. For other sites, 
\citet{Krisciunas_etal_1991} present a model of the $V$-band sky 
brightness when there is moonlight. \citet{Liu_etal_2003} also 
present a model of the brightness of moonlight as a function of 
lunar phase angle and elevation above the horizon. A much more 
advanced moonlight model, based on spectra taken at Cerro 
Paranal, Chile, is presented by \citet{Jones_etal_2013}. 
\vspace{-2em}
\section{\vspace{-1em}INSTRUMENT AND OBSERVATIONS}
\subsection{\vspace{-1em}Project goals}

\begin{figure}[!htbp]
\epsscale{0.4}
\plotone{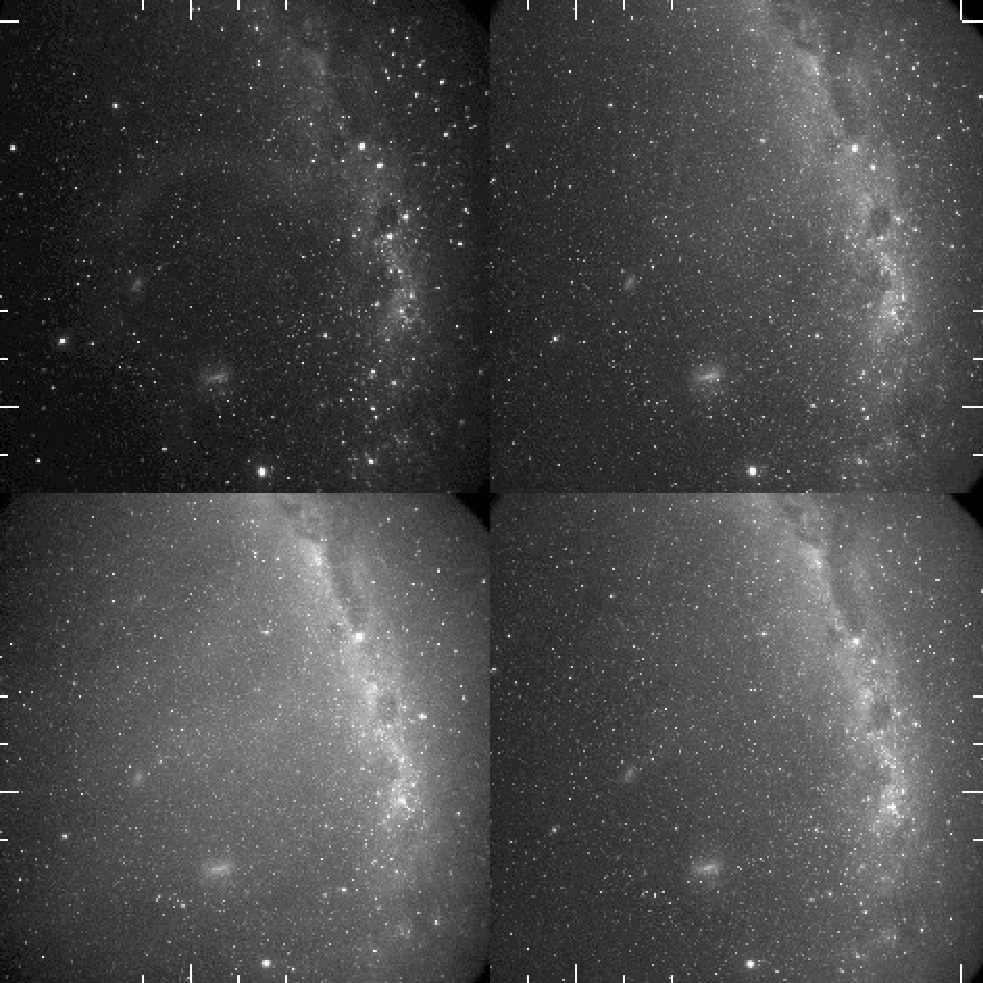}
\caption{Multi-band images obtained by GASC on 2009-06-21. 
The upper left, upper right, lower right, and lower left panels present the 
in $B$-, $V$-, $R$-, and OH-band images, respectively. The Milky Way runs from the top 
middle towards the lower right in each panel, and the LMC and SMC can be identified 
in lower left quadrant of each panel.\label{multi-band}
} 
\end{figure}

The multi-band sky brightness at Dome A, as well as the sky 
brightness among a large area of the night sky, are unknown 
quantities. The Gattini project was created to unambiguously 
measure the optical sky brightness within an incredibly large 
$90^{\circ}\times90^{\circ}$ Field-Of-View (FOV), as well as 
the cloud coverage and aurora of the winter-time sky above 
such a high-altitude Antarctic site. 
The Gattini-Dome A All-Sky Camera (GASC, \citealp{Moore_etal_2008}) 
was installed on the PLATO (PLATeau Observatory) instrument 
module, which is an automated self-powered astrophysical 
observatory deployed to Dome A \citep{Yang_etal_2009}, as part 
of the Chinese-led expedition to the highest point on the 
Antarctic plateau in January 2008. This single automated 
wide-field camera contains a suite of {Bessell} photometric 
filters ($B$, $V$, $R$) and a long-pass red filter for the 
detection and monitoring of OH emission. We have in hand one 
complete winter-time dataset (2009) from the camera that was 
returned in April 2010. 
The extremely large FOV of the GASC allows us to monitor the night 
sky brightness in the $B$, $V$, and $R$ photometric bands and the 
cloud cover beginning in the 2009 winter season at Dome A over a 
wide range ($\sim30^{\circ}$ to $0^{\circ}$) of zenith angles. 
Multi-band sky intensities measured by GASC in combination with 
spectra obtained with the NIGEL instrument \citep{Sims_etal_2010} 
will offer more comprehensive statistics on aurora and airglow. 
In addition, photometry of bright target stars in the GASC FOV with 
an unprecedented temporal window function is permitted by months of 
continual darkness during the Antarctic winter. An overview of the 
multi-band GASC FOV is shown in Fig. \ref{multi-band}. 
\vspace{-2em}
\subsection{\vspace{-1em}Dome A Camera design and assembly}
The Gattini Dome A All-sky camera was a novel low-cost pathfinder 
that ambitiously set out to measure the multi-year sky properties 
of one of the most remote and desolate sites on the planet. It was 
assembled at Caltech during 2008, and consists of a Nikon 10.5 mm 
f/2.8 GED DX fisheye lens mated to an Apogee Alta U4000 2K$\times$2K 
interline camera, and filter wheel with an assortment of photometric 
filters. The system is housed inside a heated enclosure and controlled 
by a rugged PC based supervisor system. The project cost, including 
labor for fabrication and test but excluding logistical costs, was 
approximately $50k (US$ 2008). 
The system was pointing near the SCP, without guiding or field 
rotation. The system gives a mean plate scale of approximately 
150$\arcsec$ pixel$^{-1}$, about 147$\arcsec$ pixel$^{-1}$ near 
the center of the FOV and $\sim 155\arcsec$ pixel$^{-1}$ near the 
edges of the FOV. The entire field is about $85^\circ \times 
85^\circ$.In between the lens and the camera is a 5-position 
filter wheel containing Bessell $B$, $V$, and $R$ filters 
\citep{Bessell_1990}. The remaining two slots are for a long pass 
red filter ($\lambda \textgreater$ 650 nm) for the study of airglow 
(OH) emission and an opaque mask for dark current tests.

The heat permits successful operation of the off-the-shelf camera 
as well as preventing and eradicating ice on the window surface. 
A conductive indium tin oxide coating was employed on the window 
surface as a method of de-icing. However, it was found that when 
operational this was not sufficient to remove ice that was 
deposited in large amounts due to the surface wind. 
Some images showed very few sources over a small consecutive 
period of time, ranging from hours up to $\sim$3 days. It is not 
possible to determine whether the lack of astronomical sources was 
caused by ice and frost formed on the cover window, or because of 
the atmosphere. The internal heating was sufficient to keep the 
window ice-free during the rest of the winter period. 

The experiment was controlled by a low-power computer in an 
electronics rack inside the PLATO module. The continuous observation 
was operationally simple and repeated this sequence: 
$B$-band (100 s and 30 s), 
$V$-band (100 s and 30 s), $R$-band (100 s and 30 s), $OH$-band as a 
long pass red filter (100 s and 30 s), 100 s dark, and bias frame. 
During periods of bright twilight at the beginning and the end of the 
2009 winter season, the camera adjusted its exposure time to adapt to 
the sky counts, or truncated integrations to prevent over-exposure.
\vspace{-2em}
\subsection{\vspace{-1em}The 2009 Data Set}
The full dataset contains approximately 160,000 images obtained from 
2009-04-18 to 2009-10-10. Unfortunately, the images obtained 
before 2009-05-19 were affected by snow over more than one-third of 
the FOV. Considering the unknown transmission and reflection caused 
by the anisotropic snow coverage, those images were discarded. 
Additionally, the sky became continuously bright due to the {Sun} 
after 2009-09-18, so we excluded from the data reduction all the 
images obtained after that date. 
{Images obtained prior to that date, however, with Sun's elevation angle 
greater than $-10^{\circ}$ and median count rate above a certain threshold, 
have also been excluded.} Images with a 100 second exposure 
time have been used to inspect and calibrate the sky brightness. 
Approximately 11925 frames were obtained in each photometric band 
for each exposure time between 2009-05-19 and 2009-09-18 {($\sim$ 123 days)}, 
resulting in a total of 331.25 hours of 100 s exposures in each 
filter. {The typical cadence for the $B$, $V$, and $R$ band 100 s 
exposures sequences gives $\sim$737 s.}
\vspace{-2em}
\subsection{\vspace{-1em}Instrumental Effects{\label{sec_inst}}}
{GASC was set out to 
measure the multi-year sky properties of one of the most remote 
and desolate sites on the planet.} Given the nature of the wide 
field design, combined with an interline CCD and no sidereal 
tracking system, non-negligible effects needed to be modeled to 
process the data effectively. These effects are summarized as 
follows. 
\vspace{-2em}
\subsubsection{\vspace{-1em}Absence of a Sidereal Tracking System}
The camera was mounted in a heated enclosure, with fixed pointing 
in the direction of the SCP. The absence of a mechanical tracking 
system, together with an extremely large FOV, produces stellar 
images that exhibit a different Point Spread Function (PSF) at 
different positions throughout the FOV. Over the course of the 
100 s exposures the stars produced elongated circular tracks 
owing to the Earth's rotation. This effect is most obvious for 
stars furthest from the exact location of the SCP, which was 
close to the center of the GASC FOV. The observed largest 
elongations are $\sim$6 pixels in the $X$ direction and $\sim$6 
pixels in the $Y$ direction for each 100s exposure frames. 
\vspace{-2em}
\subsubsection{\vspace{-1em}Angle between the Optical Axis and the South Celestial Pole}
\begin{figure}[!htbp]
\epsscale{0.55}
\plotone{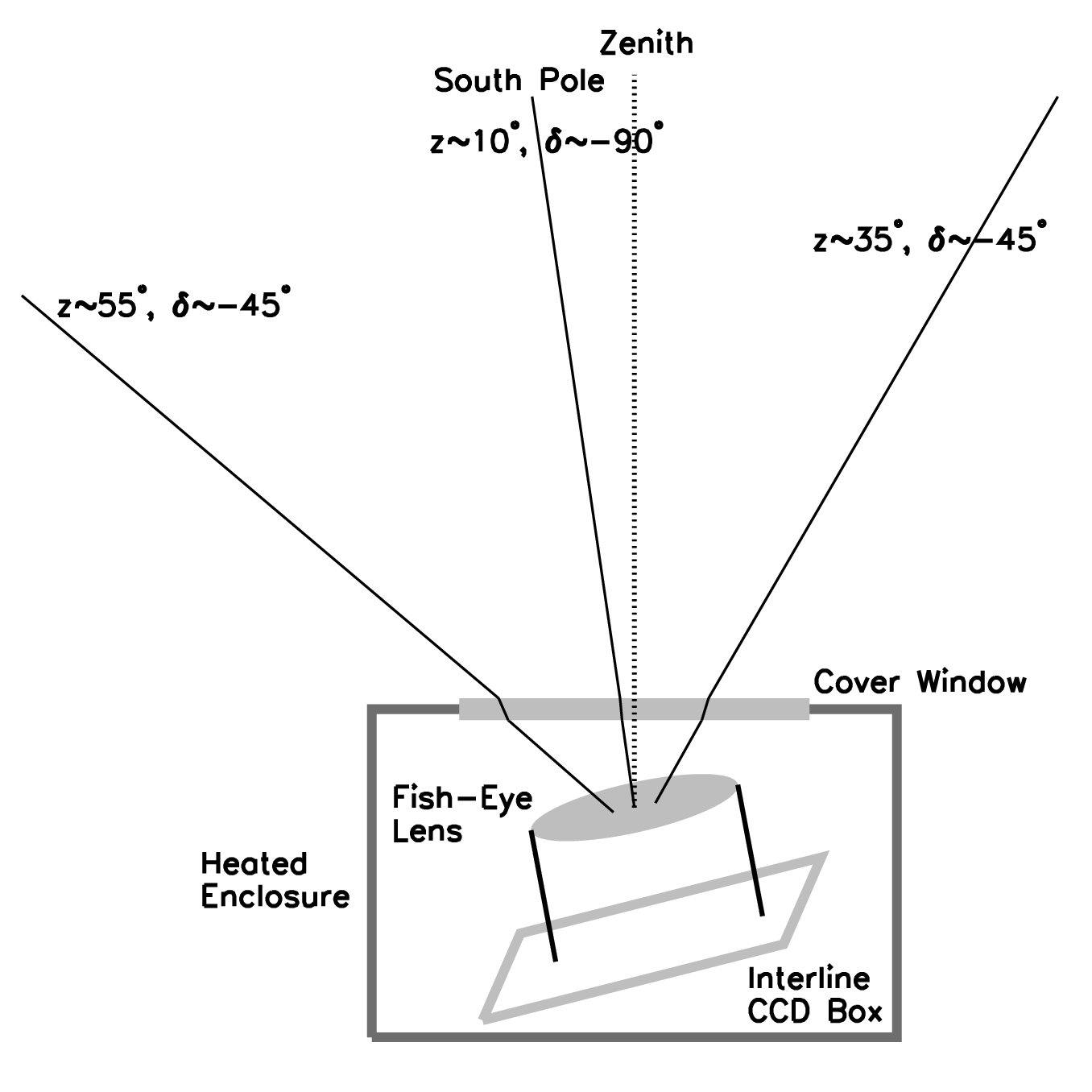}
\vspace{-1em}
\caption{Schematic diagram showing the set-up of GASC.
\label{schem_window}} 
\end{figure}

The optical axis of the camera was closely aligned with the SCP, 
while the horizontal axis of the cover window of the heated 
enclosure was fixed to be aligned with the zenith. Because stars 
obtain different zenith angles as a result of the travel along 
declination circles, both the optical path length through the 
material of the cover window and the thickness of the Earth's 
atmosphere changes with stellar azimuths and elevation angles, 
introducing a periodic, asymmetrical variation of the stellar 
light curves, even for stars of constant brightness. 
Additionally, since the GASC camera was pointing near the SCP, 
while the cover window was pointing at the zenith, the 
$\sim 10^\circ$ offset between the SCP and zenith introduced 
transmission differences as the light passed through the cover 
window. A schematic of this set-up is presented in Figure 
\ref{schem_window}. 
\vspace{-2em}
\subsubsection{\vspace{-1em}Vignetting}
In large field astronomical images, as well as photography and 
optics, vignetting causes a reduction of flux at the periphery 
compared to the image center. 
In the optical design of GASC, Vigentting was necessary to 
minimize the scattered light from the {Moon} when it is above the 
horizon. However, vignetting also significantly reduces both the 
flux from the stars and from the sky background, especially at 
the edges of the GASC FOV. 
\vspace{-2em}
\subsubsection{\vspace{-1em}Interline Transfer Sensor}
The detector situated behind the multi-band filter wheel is a 
2K$\times$2K interline transfer CCD. It has a parallel register 
that has been subdivided into two stripes to create opaque storage 
register fits between each pair of columns of pixels. These opaque 
masks occupy a large portion of the area of the CCD. Although 
micro-lenses have been annealed to the CCD that focus light from a 
large area down to the photo-diode, when light beams are incident 
at large angles, the micro-lens array will fail to direct all the 
photons directly down to the photo-diode. For this reason, the CCD 
is less sensitive to some incident directions of light. As the 
stars move around the SCP this effect will cause periodic 
fluctuations in the resulting light curves. The amplitude of those 
variations is strongly correlated with a star's angular distance 
from the SCP. For the GASC optical system, this effect can 
reach $\sim0.2$ magnitude. Further test and analysis will be 
presented in Section \ref{section_palomar}. 

\vspace{-1em}
\section{\vspace{-1em}Data Reduction}
\begin{figure}
\epsscale{0.8}
\plotone{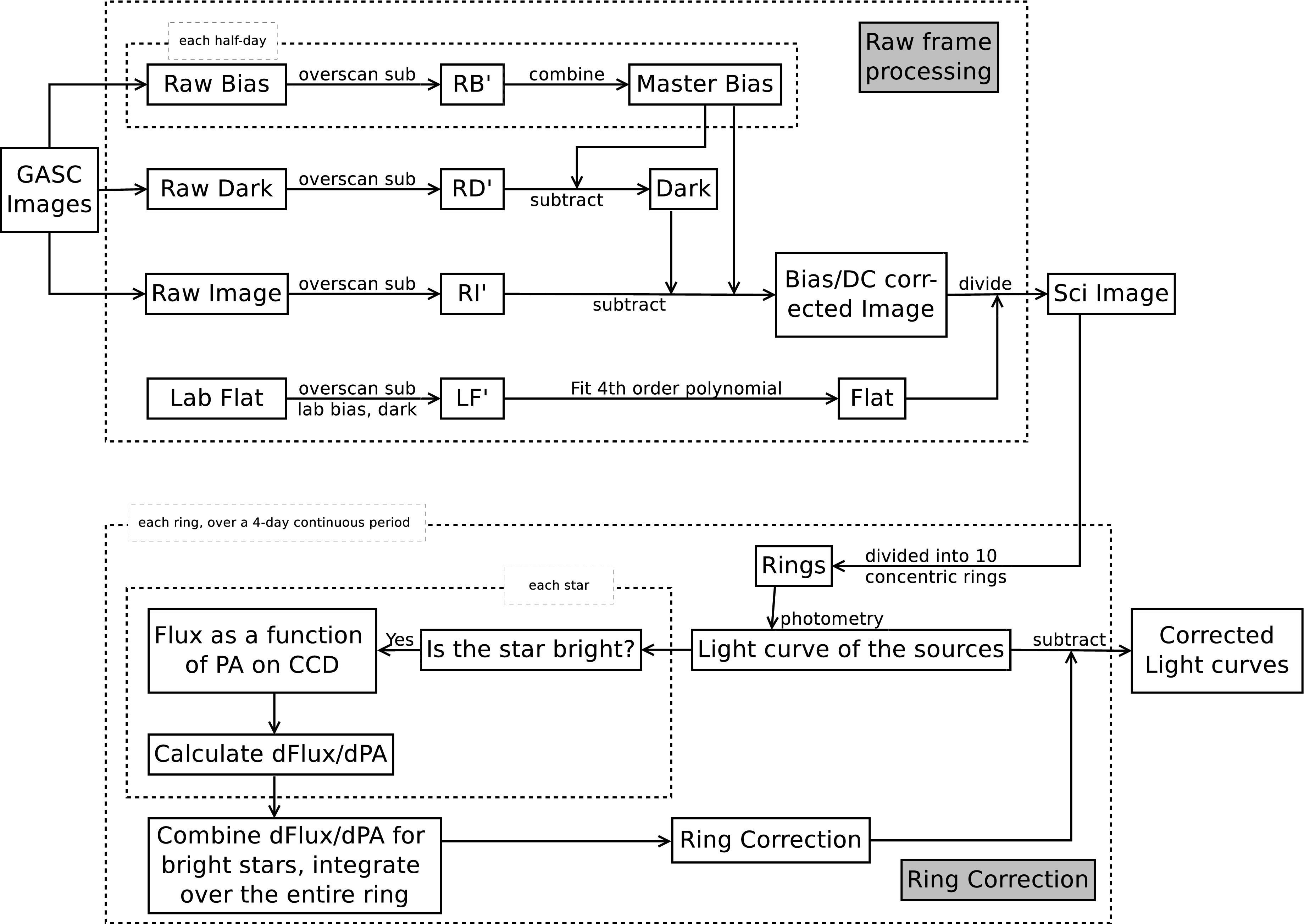}
\vspace{-1em}
\caption{Flow chart showing the customized GASC data reduction pipeline.
\label{flowchart}} 
\end{figure}

GASC has a FOV approximately $85^\circ \times 85^\circ$, and the 
absence of a mechanical system for tracking will lead to star trails 
on the CCD over the course of the exposures. The instrument is 
fixed in orientation and stars sweep out {circular arcs} centered on the 
South Pole every sidereal day. The illumination response of the 
GASC across the large FOV is highly varying, in cases up to 30\% 
variation from the center to the edge of the field, due to inherent 
qualities of the fish-eye lens and due to mechanical baffling 
introduced to minimize the scattering of light due to the moon. In 
addition, there are sidereal variations {on} the order of 0.2 
magnitudes, due to instrumental effects expanded in Sections 
\ref{sec_inst}. A custom data reduction pipeline is comprised of a set 
of routines written in IDL that processes the $\sim$11925 raw sky frames 
for each filter band and produces calibrated sky brightness 
measurements. The pipeline by necessity also produces calibrated 
light curves of all the stars brighter than $\sim$7.5 in $V$. An 
overview of the essential steps is presented in Figure 
\ref{flowchart}. Each step is detailed in the respective sub-section 
below. 
\vspace{-2em}
\subsection{\vspace{-1em}Pre-reduction}
The overscan region of each frame was subtracted to remove the 
consequences of any voltage variations. In each half-day period of 
observation, a ``master bias frame'' was made by combining single 
overscan-subtracted bias frames. For each half-day period of 
observation, this ``master bias frame'' has been subtracted from the 
data frames to remove the internal bias structure across the chip. 
The internal temperature variations within the heated enclosure may 
lead to implied (and artificial) variations of the sky brightness as 
well as the photometry of bright targets. We tested the possibility 
that the enclosure temperature and CCD temperature affect the 
photometric magnitudes by calculating any possible cross-correlations 
between the enclosure temperature, the CCD temperature, together with 
typical light curves for bright stars in the GASC FOV during the 
entire 2009 winter season. No correlations between any pairs of those 
factors have been identified, indicating a stable work state of GASC 
during the 2009 winter season and a reasonable bias subtraction 
technique. 

Acquisition of usable sky flats for this type of system is difficult to 
perform on the sky, due to the non-tracking capability of the system and 
the sheer size of the FOV. 
We measured the flat field illumination properties of the GASC with a 
uniform illumination screen after the system was returned to Caltech from 
the Dome A. 
A multi-band lab flat shows that the optical center of the lens is, 
fortuitously, coincident with the SCP. For each photometric bandpass, 
a fourth order polynomial has been directly applied to fit 
the lab flat. The lab flat was used as a method to remove global 
transmission variations across the field, whereas, pixel-to-pixel 
variations were removed by compiling a sky reference flat. The 
pixel-to-pixel variations turn out to be negligible (less than 
$\sim0.3\%$) when compared to the photometric accuracy GASC is 
able to achieve.
However, it was not able to remove the remaining $\lesssim$0.2 mag
variations that were removed by the ``ring correction'' technique, which 
will be discussed in Section \ref{sec_ring}. 
\vspace{-2em}
\subsection{\vspace{-1em}Image Profiles and Astrometry}
\begin{figure}[!htbp]
\epsscale{0.75}
\plotone{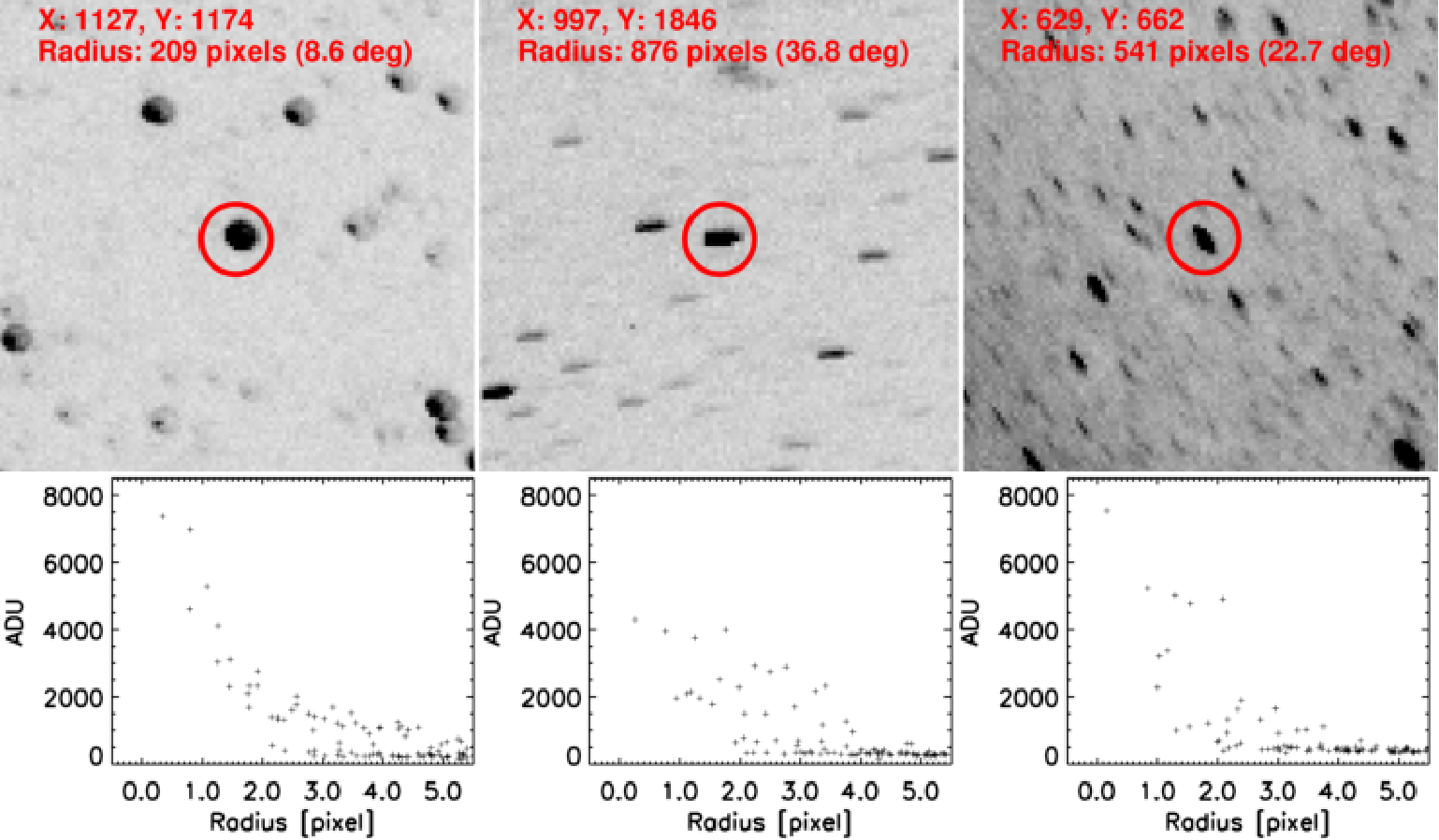}
\vspace{-1em}
\caption{Typical profiles of stars at different distances to the SCP. 
\label{psf_look}} 
\end{figure}

The {\sc daofind} and {\sc apphot} packages within 
{\sc iraf}\footnote[1]{{\sc iraf} is distributed by the National 
Optical Astronomy Observatories, which is operated by the Association 
of Universities for Research in Astronomy, Inc., under cooperative 
agreement with the National Science Foundation (NSF).} were used to 
detect and perform photometry on approximately 2600 bright stars in 
the GASC FOV, most of which are between 3.5 to 7.5 magnitude in $V$. 
Without tracking, stars trail along concentric rings around the SCP 
and present elongated, curved PSFs on each frame. 
Figure \ref{psf_look} presents the typical profiles of stars at 
different distances to the SCP. 

The astrometry routine adopted in the GASC data reduction pipeline 
makes use of the almost-polar location of the instrument. 
We derotated the physical coordinates of the 
sources in each image relative to known reference images. 

\begin{figure}[!htbp]
\epsscale{0.85}
\plotone{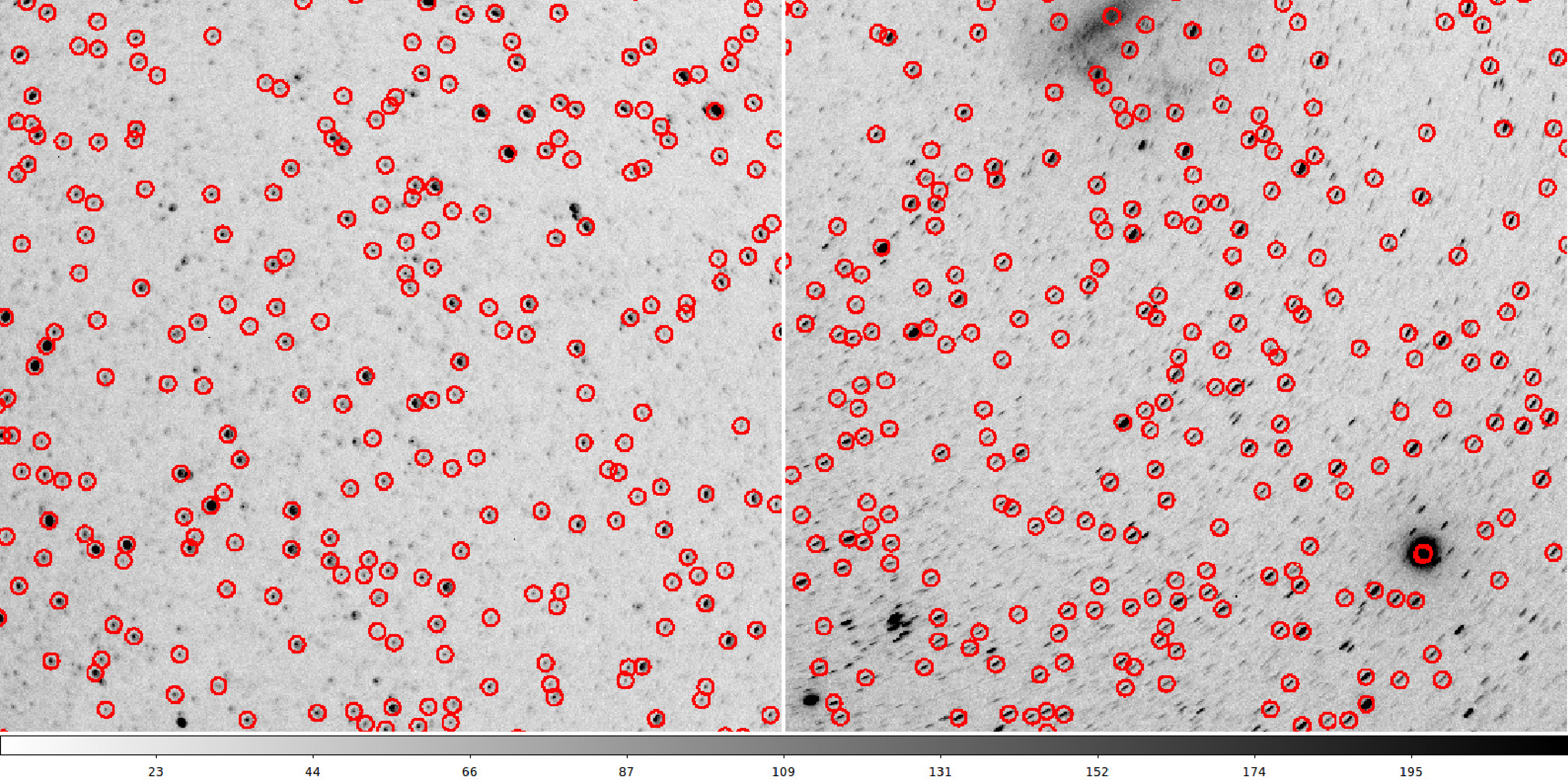}
\vspace{-1em}
\caption{The stellar field in the GASC FOV obtained on 2009-06-22. The left panel 
shows the central FOV and the right panel shows one corner of the FOV. Sources 
chosen to perform aperture photometry have been circled by $r = 4$ pixel apertures. 
The images were taken in defocussed mode to account for the huge pixel scale. 
The right panel shows significant star tracks near the corner of the FOV due to 
the Earth's rotation.
\label{center-corner}} 
\end{figure}

To reduce the uncertainty caused by distortion 
and increase the accuracy of matching, as reference frames we 
selected 20 high quality frames equally spaced in time over one 
entire rotation cycle (i.e. one sidereal day). Given the time of 
exposure of any other frames, all stellar coordinates can be 
obtained by rotating those 20 templates within $\pm$9 degrees. 
This provided a time economical solution for performing the astrometry 
required by the GASC science goals on the $\sim$36,000 sample images. 
An overview of the GASC FOV and field stars used to perform 
aperture photometry is shown in Fig. \ref{center-corner}.

\vspace{-2em}
\subsection{\vspace{-1em}Ring Correction \label{sec_ring}}
Due to the combined effect of the presence of the cover window and 
the different response of the interline transfer CCD to different 
incident angles, light curves for $\sim2600$ stars imaged in the 
GASC FOV show asymmetrical sidereal fluctuations. The amplitude of 
this variation grows as the distance of stars to the SCP increases. 
We looked at the behavior of bright, isolated stars which sweep out 
concentric rings in the GASC FOV. As the ``standard stars'' have 
higher S/N, a weighted combination of their light curves {gives} us 
feedback {on} the entire optical system. This feedback, however, also 
applies for all the other stars with a lower S/N.

\begin{figure}[!htbp]
\epsscale{0.4}
\plotone{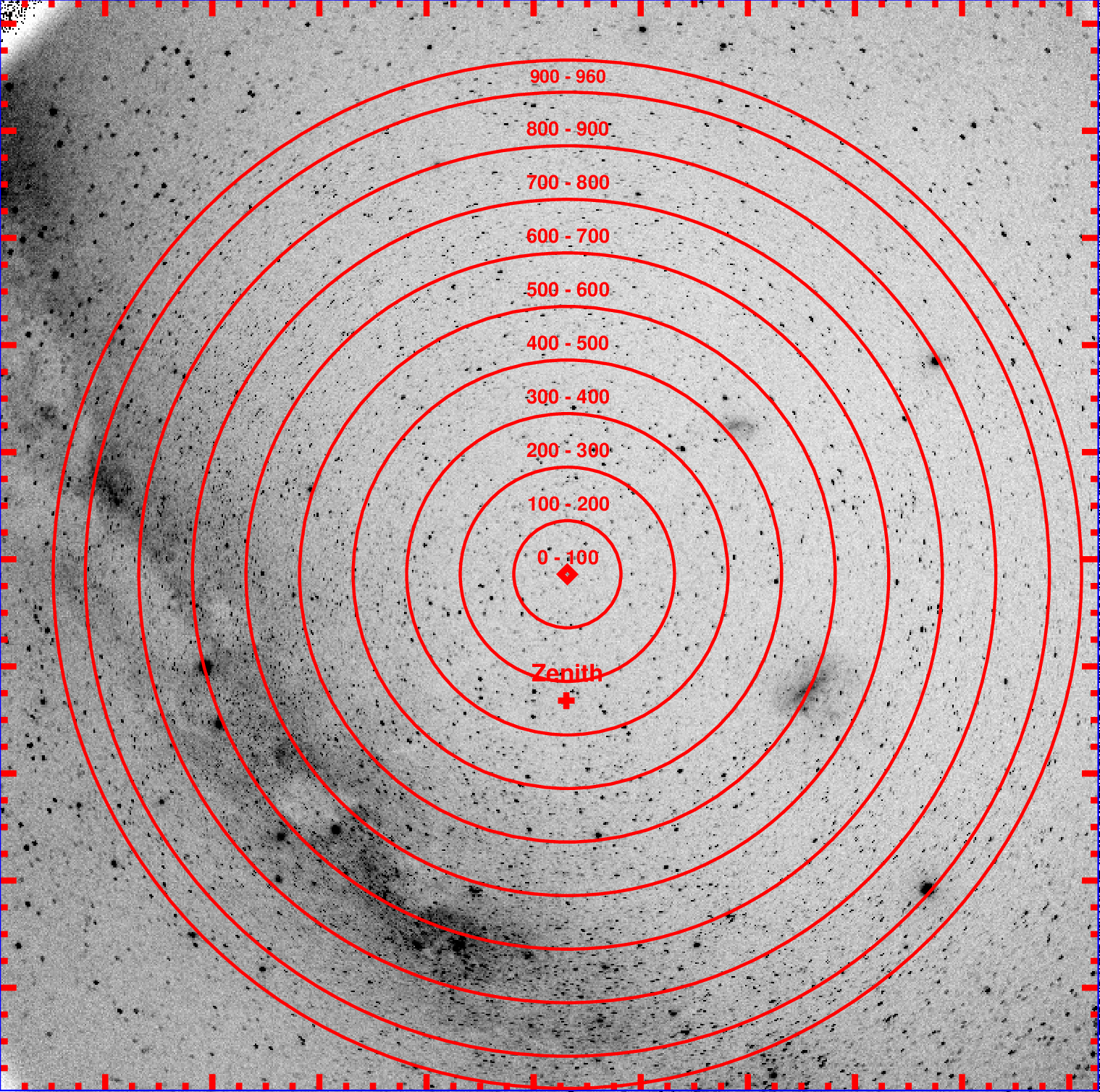}
\vspace{-1em}
\caption{Concentric rings dividing the GASC FOV. 
The $+$ marks the physical position of {the zenith} on the GASC FOV.
\label{ring_look}} 
\end{figure}

Here we introduce a ``ring correction'' to remove the residual 
instrumental effects, to the order of $\pm$0.2 mag in the raw 
photometry. The methodology is to consider 
the features of the light curves for {bright} stars that have similar 
distances from the SCP, as they sweep out paths along the same 
ring with different hour angles but similar declination.  
The systematic light curve features do not change drastically as 
slightly different radii. 
The GASC FOV has been subdivided into 10 concentric rings, each 
with a width of 100 pixels with an exception of 60 pixels for the 
outermost ring. Figure \ref{ring_look} shows the concentric rings 
dividing the GASC FOV. 
Within each ring we investigated the behavior of standard stars 
{which are non-variable stars with $V\sim 3.5 - 5.5$}, 
mapped the gradient of its flux variation over different position 
angles ($PA$) on the CCD chip relative to the SCP, i.e., 
$\mathrm{d} Flux / \mathrm{d} PA$. Then, we combined the gradients 
calculated from each standard star at each $PA$ over a continuous 
run of observations under good weather conditions, and applied a 
spline interpolation to obtain a gradient map over that ring. We 
then integrate over the $PA$ and convert the integrated flux into 
a magnitude. This produces a phase diagram of magnitude variations 
within each ring, representing systematic behavior of the stars as 
they trail along certain rings of the CCD chip. 

\begin{figure}[!htbp]
\epsscale{0.90}
\plotone{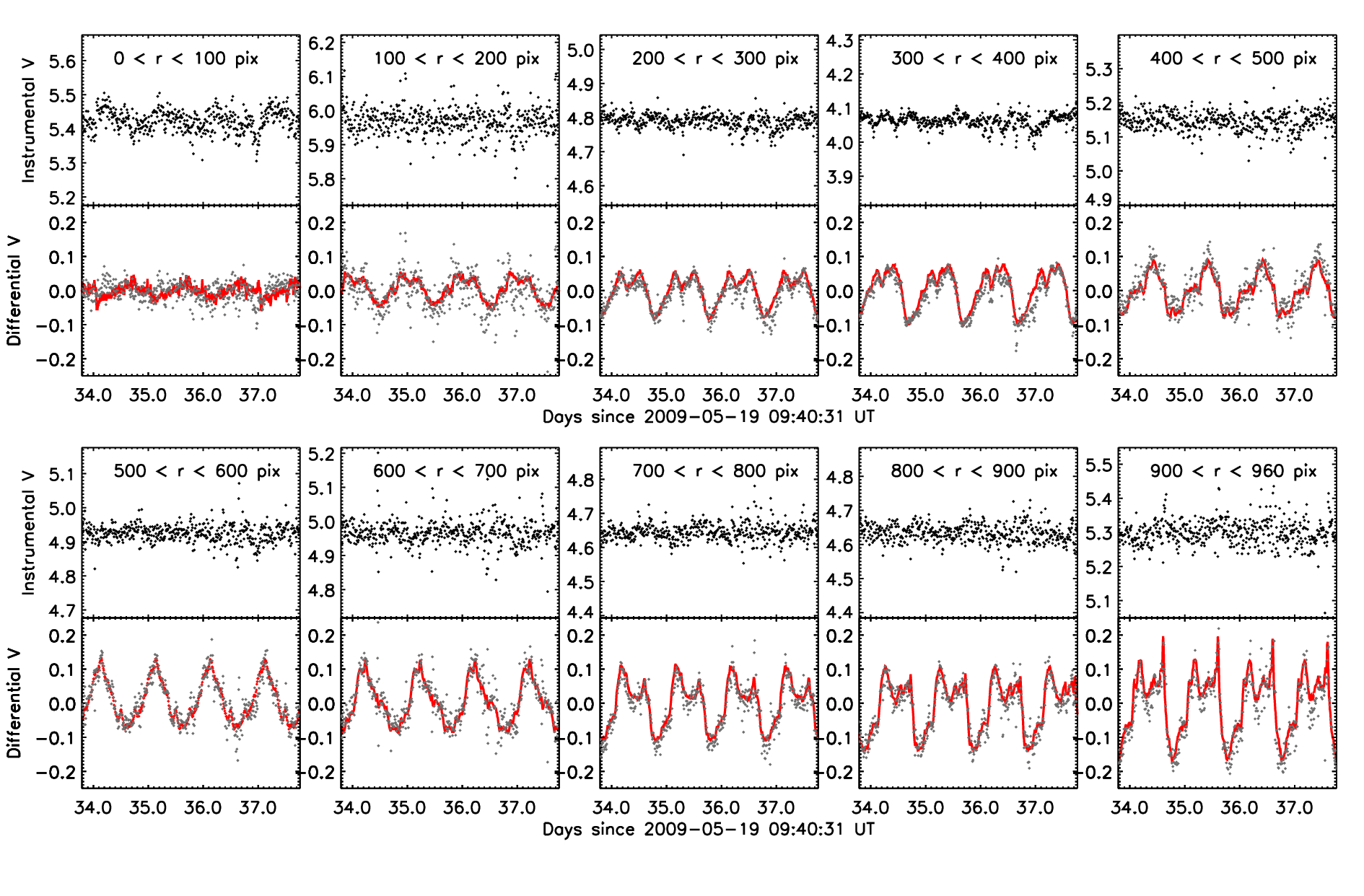}
\vspace{-1em}
\caption{The ``ring corrections'' for $V$-band light curves for 10 different 
annuli are shown as 10 sub-figures. Each panel represents an annulus 
width of 100 pixels in radius. The upper sub-panels 
represent the output light curves after applying the ring corrections. In 
the lower sub-panels, the gray dots represent the input light curves before 
applying the corrections, and the red symbols represent the models of 
corrections within each corresponding radius range.
\label{ring-corrs}} 
\end{figure}

\begin{figure}[!htbp]
\epsscale{0.55}
\plotone{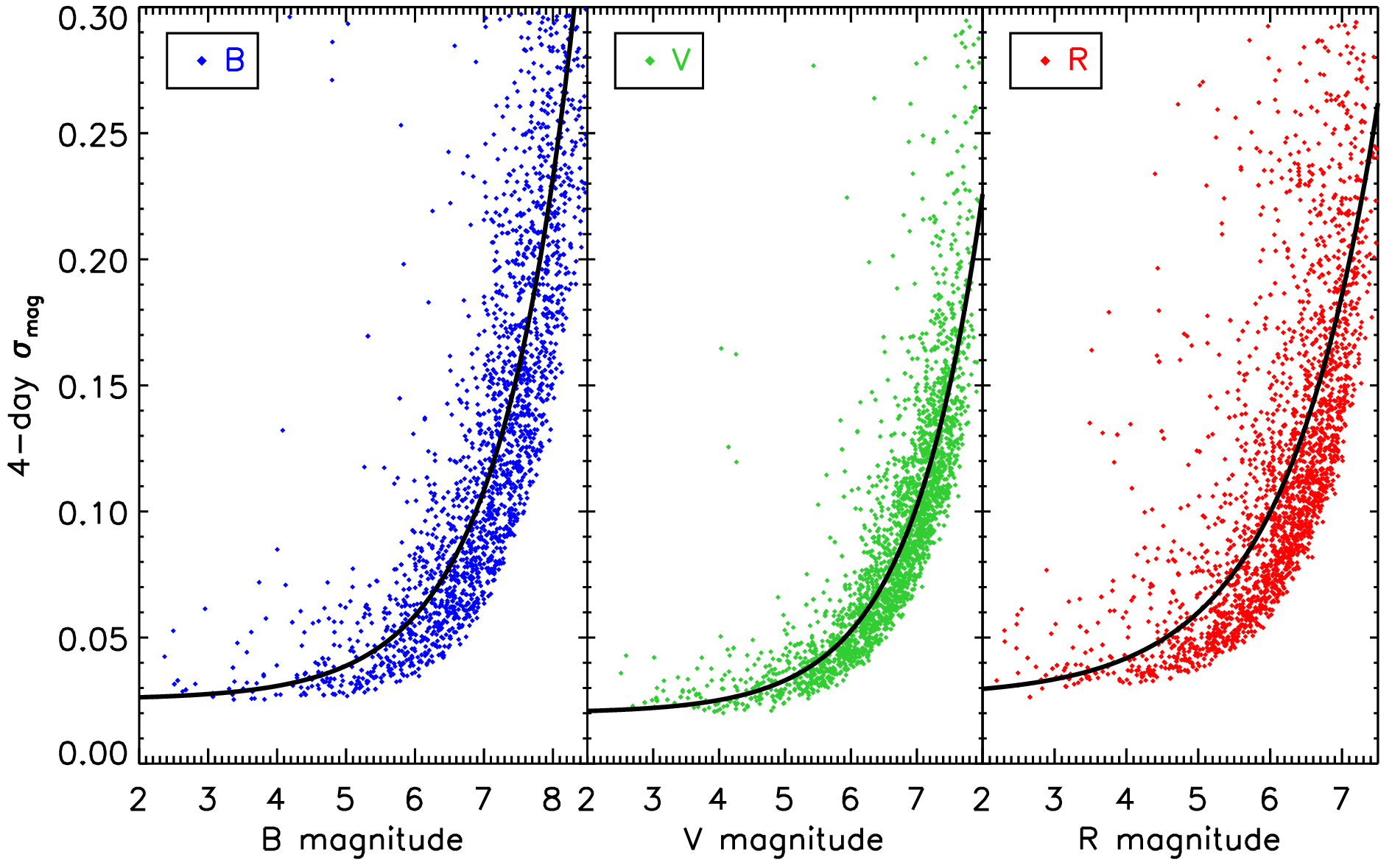}
\vspace{-1em}
\caption{Photometric errors vs. stellar brightness. From left to right 
we show the photometric accuracy in the Bessell $B$-, $V$-, $R$-bands, 
respectively, after applying the ``ring corrections.'' The photometric 
uncertainties were calculated from data obtained on 4 consecutive days. 
\label{sig-mag}} 
\end{figure}

We refer to this procedure as the ``ring correction''. 
The light curve corrections for all the other stars can be obtained 
by subtracting the ``ring correction'' after proper time phase 
matching. The ``ring corrections'' have been built based on a 4-day 
continuous run of high-quality data obtained from 04:25 UT on 2009-06-22 
to 03:47 UT on 2009-06-26. This has been applied 
successfully to the data obtained during the entire season.
The ``ring corrections'' for typical stars within each one of the ten 
rings are shown in Fig. \ref{ring-corrs}, and they work well for most 
of the cases. Additionally, the $\sigma-$ magnitude diagram, after 
applying both the pseudo-star correction and the ring corrections is 
shown in Fig. \ref{sig-mag}. For instance, we obtained $\sim3\%$ 
photometric accuracy for stars with apparent magnitude 
$V \approx$ 5.5. 
In summary, with the ring correction procedure completed, 
the light curves have been corrected for instrumental effects that cause 
intensity {variations} across the field and as a function of time. 
\vspace{-2em}
\subsection{\vspace{-1em}Calibration for Sky Brightness}
\subsubsection{\vspace{-1em}Determination of Catalog Magnitude}
\begin{figure}[!htbp]
\epsscale{0.55}
\plotone{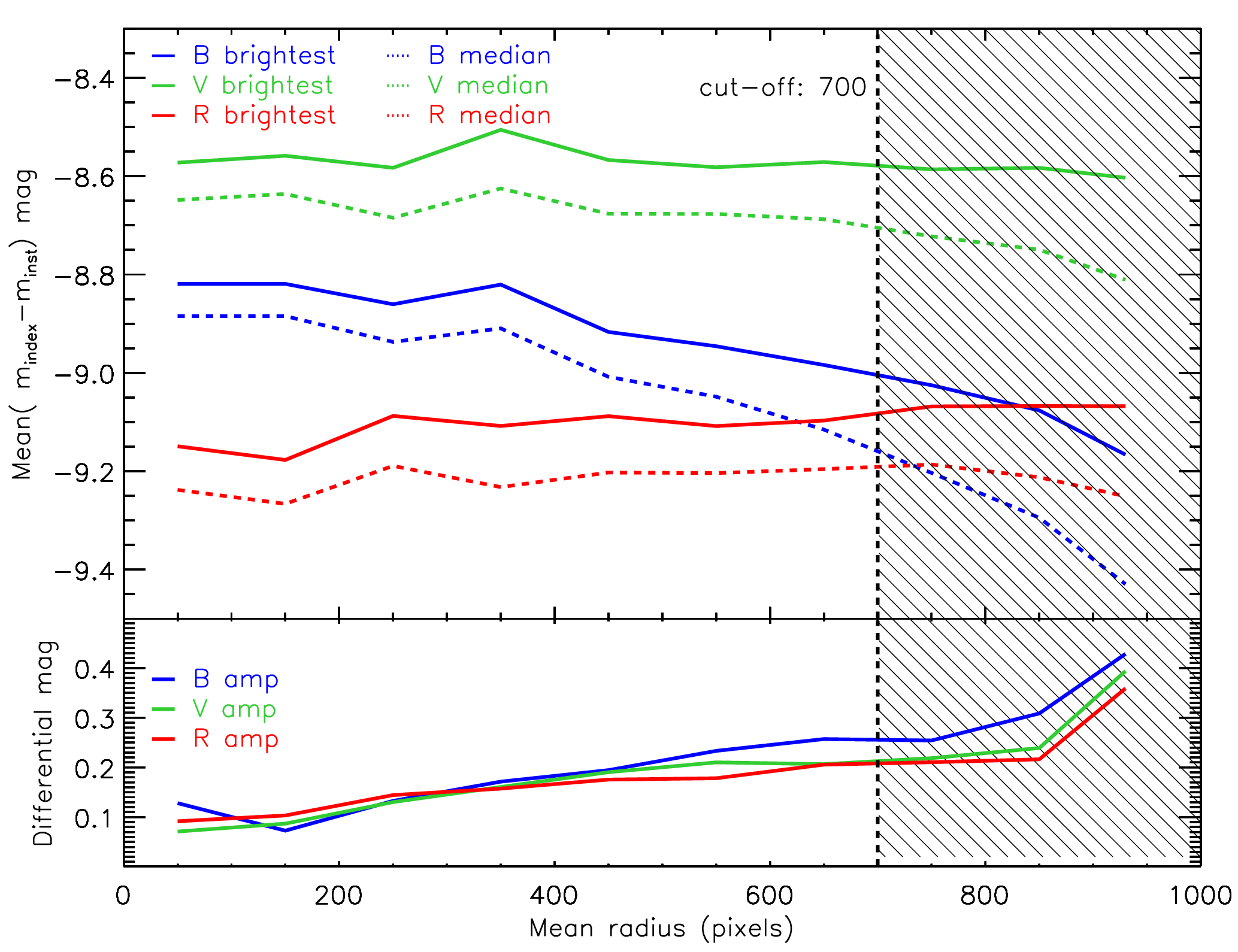}
\vspace{-1em}
\caption{A radius-magnitude 
offset diagram for the ``ring correction'' for different radii. The offset 
between standard stars' catalog magnitude and instrumental magnitude has been 
calculated based on two different considerations of instrumental magnitude. The 
results are based on the median values of all the standard stars' 
brightest (represented by solid lines) and median (represented by dashed lines)
magnitudes during a sidereal day. 
The lower panel shows radius-amplitude diagrams for the ``ring correction'' 
in different annuli. 
A significant increase in amplitude occurs if the radius is 
increased from 700 pixels to 800 pixels. A vertical dashed line and the shaded 
region indicate the 700 pixel radius cut-off for stars to be used for 
calibration.
\label{diff-radius}} 
\end{figure}

This step of the GASC data reduction pipeline converts instrumental 
magnitudes to catalog magnitudes. Furthermore, the sky brightness 
can be determined by applying this offset to the GASC measured sky 
flux. If we define the ``radius'' of each star as its distance from 
the SCP (pixel coordinates $X$ = 1063, $Y$ = 972), we find that the 
amplitude of the daily fluctuation in a star's light curve depends on 
(1) its radius, (2) observing bandpass, as shown in Fig. \ref{diff-radius}, 
together with (3) the mean value of the difference 
between the standard star's catalog and instrumental magnitude. 
As radius increases, the more stable the mean difference between a 
standard star's catalog and instrumental magnitude, the less 
affected the standard star's flux is due to instrumental effects. 

The upper panel in Fig. \ref{diff-radius} shows 
a stable trend of the mean difference between standard stars' catalog 
and instrumental magnitudes in the $V$- and $R$-bands, which means 
that as stars travel around in GASC FOV, though the distance to the 
SCP varies for different stars, it is still reasonable to treat the 
brightest magnitude in one cycle as the true instrumental magnitude of 
that star. 
Instrumental effects become more significant near the edges of the FOV. 
Strong geometrical distortions, as well as the large incident angle 
near the edges of the FOV, will cause an unexpected and non-negligible 
reduction of the flux transmitted through the optical system. 
Giving special consideration to the case of the $B$- and $V$-bands, 
we set a cutoff radius of 700 pixels, corresponding to 
$\sim30^{\circ}$ from the SCP, and we use all the standard stars 
within this radius to calibrate the sky brightness. 

We rely on the linearity of the CCD and minimize the $\chi^2$ value 
of the fit using the offset between the standard stars' photometric 
magnitudes and their catalog magnitudes. We consider data only within 
700 pixels of the SCP and weight by the area of each ring. This gives 
us our multi-band sky brightness measures at Dome A calibrated by the 
standard stars. Once GASC was shipped back to Caltech, we performed 
tests at Palomar Observatory. Table \ref{cal_models} gives the $BVR$ 
photometric offsets from instrumental to calibrated values. The details 
of the Palomar GASC test are discussed in the next section. The offset 
in the constant term between Palomar and Dome A calibration model was 
due to the absence of the cover window in the Palomar test and the 
different exposure times between two observation epochs.
\vspace{-2em}
\subsubsection{\vspace{-1em}Determination of Photometric conditions\label{sec_pseudostar}}
Variations of global transparency, including weather changes, possibly 
snow and frost formed in front of the enclosure's cover window, will 
dramatically affect many quantities in measuring sky brightness, the 
fraction of the sky covered by cloud, as well as photometry of bright 
sources. This global effect can be subtracted off by introducing a 
``pseudo-star'', with a count rate $f^p$ and instrumental magnitude 
$m^p = -2.5 \times \mathrm{log}_{10} f^p$, which has been constructed 
from the observed counts of $~2600$ target stars in each frame 
according to:
\begin{equation}
f^{p}_{i}  = \; \sum \frac {f_{i,j}} {(\sigma^{ring}_{j})^2  + (\sigma_{i,j})^2}  \; , 
\ \  m^{p} = -2.5 \times \mathrm{log}_{10} f^p + ZP^{P}
\label{eqn_pseudo_star}
\end{equation}
where $i$ is the frame number in the 
observing sequence and $j$ is the star number in each frame. 
$\sigma^{ring}_{j}$ gives the standard deviation of the residuals 
in counts for the $j^{th}$ star after the ring correction during 
the 4-day continuous run of high-quality data obtained from 
04:25 UT on June 22 to 03:47 UT on June 26, 
$\sigma_{i,j}$ gives the measured photometric error for the 
$j^{th}$ star in the $i^{th}$ observation, $ZP^{P}$ is the 
zero point for instrumental magnitude and assigned to be 25. 
We subtract $m^p$ from the rough photometric results 
to remove the global variations in the entire GASC FOV. 
Furthermore, the variation of the pseudo-star can be an 
indicator of transparency variations and further used to 
estimate the cloud coverage. A more detailed discussion will 
be presented in the following sections.
\vspace{-2em}
\subsubsection{\vspace{-1em}GASC Test at Palomar Observatory\label{section_palomar}}
To test the quality of GASC measurements and the calibration of sky 
brightness, another experiment intended to measure and calibrate the 
sky brightness at an astronomical site was implemented at Palomar 
Mountain Observatory. The sky at Palomar during a moonless night is 
sufficiently dark to check the Dome A measurements. The Palomar Night 
Sky Brightness Monitor (NSBM)
\footnote[2]{http://www.sao.arizona.edu/FLWO/SBM/SBMreport\_McKenna\_Apr08.pdf}
allows a real-time comparison between the night sky brightness 
measured by the two different instruments.
The Palomar NSBM consists of two units deployed at Palomar Observatory. 
A remote photometer head and a base station receive data from the remote 
head via a wireless spread-spectrum transceiver pair. The remote head has 
two photometers that sample areas of the sky $\sim5.6^{\circ}$ in diameter 
at two elevation angles. The photodetectors used to measure the sky 
brightness receive filtered light to define a spectral response centered 
in the visual range, with a strong cutoff in the near infrared. 

\begin{figure}[!htbp]
\epsscale{0.5}
\plotone{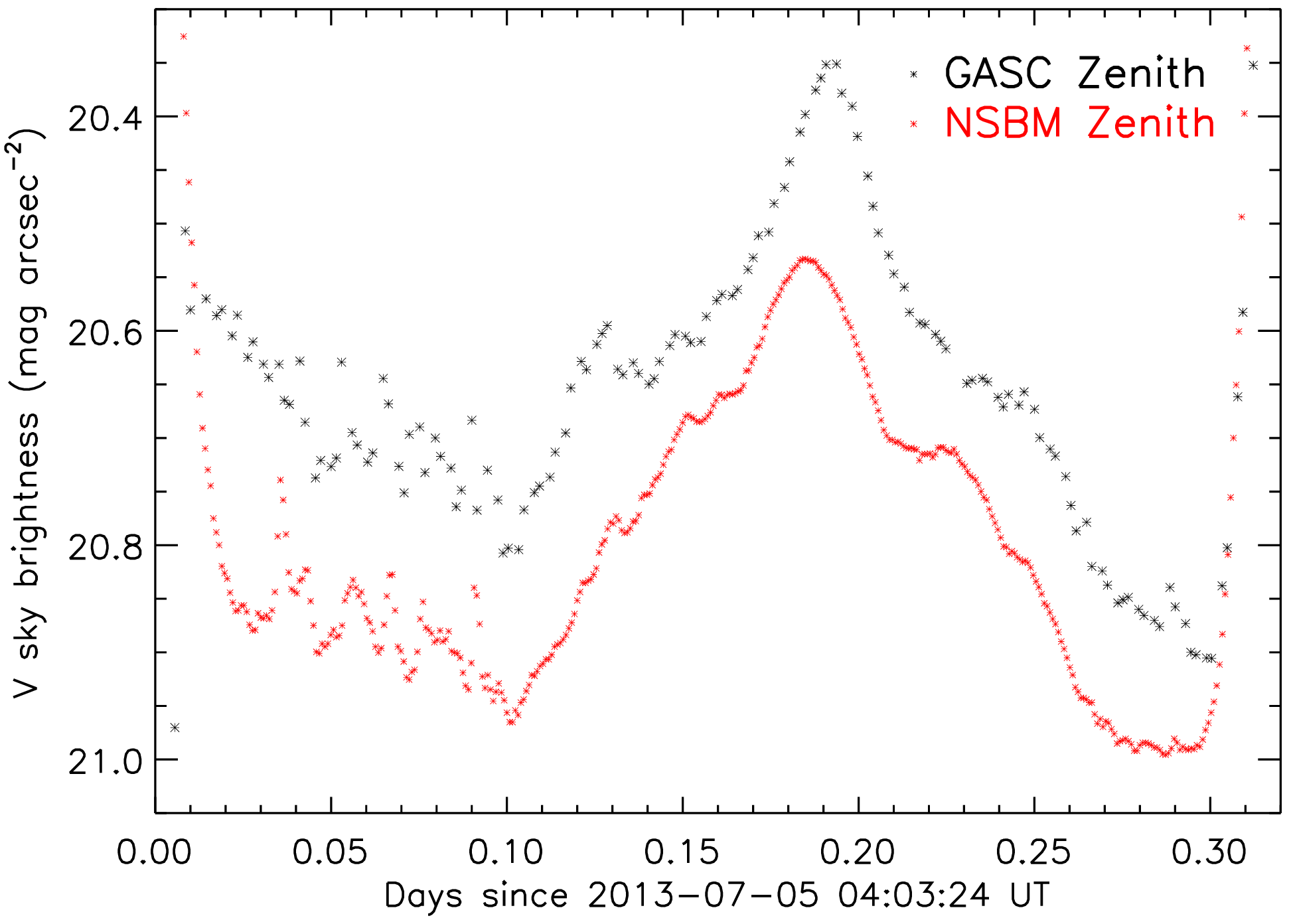}
\vspace{-1em}
\caption{Palomar night sky brightness measured and calibrated 
by NSBM (red dots) and GASC (black dots) on UT 07-05-2013. 
\label{palomar-NSB}} 
\end{figure}

One unit of the NSBM uses a 1.5-cm diameter photodetector, which measures 
the brightness of the sky $\sim5.6^\circ$ in diameter at the zenith. Without 
rejecting stellar contaminants the mean value for this region is taken to 
represent the night sky brightness. The output data from the NSBM consists of 
the measured frequency and ambient temperature of each sensor. The sky 
brightness is calculated as\footnote[3]{Zenith readings are available at: 
http://www.palomar.caltech.edu:8000/maintenance/darksky/index.tcl}: 
\begin{equation}
{ Zenith \; magnitude} \; = \; -2.5 \; \mathrm{ log}_{10}({ Zenith \; reading} 
- 0.012) \; + \; ZP \; .
\label{eqn_nsbm}
\end{equation}
The detector output frequency (in Hertz) constitutes the raw data, 
as the NSBM uses a light to frequency converter. The dark frequency 
to be subtracted for the zenith is 0.012 Hz. The zero point of the 
NSBM system adjusted to the National Parks System from one night's 
data (4 July 2013), is 19.41 mag arcsec$^{-2}$ and for a band 
comparable to the Johnson $V$-band is 18.89. Fig. \ref{palomar-NSB} 
shows the time variations of the sky brightness measured by NSBM and 
GASC. The sky brightness measured by two different instruments, with 
two completely different calibration methods agrees overall to 
$\sim0.12$ mag arcsec$^{-2}$.

A separate test was conducted at Palomar to show that the camera 
orientation, specifically the azimuth angle of the camera, results 
in a variation in the magnitudes of bright stars. This test used 
exposures taken very close to one another in time. The results of 
this test confirmed the variations we see in the original data. 
\vspace{-1em}
\section{\vspace{-1em}Results and Discussion}
\subsection{\vspace{-1em}Sources of Sky Brightness}
Artificial light pollution is essentially nonexistent at Dome A, Antarctica. The 
main contribution to the sky background is usually from the atmospheric scattering 
of the light from the Sun and the Moon. 
At Dome A (80$^{\circ}$ 22$\arcmin$ S, 77$^{\circ}$ 21$\arcmin$ E) there is some twilight 
time even on the first day of the southern winter, as the Sun is roughly 
$-13.8^{\circ}$ below the horizon at local noontime. 
The closer the Sun is to 
the horizon at local noontime on other days of the year, the greater will be the 
variation of the sky brightness, even on days when the Sun does not rise and set.

Airglow persistently provides photon emission and gives the 
dominant component of the optical and near-IR night sky brightness 
\citep{Benn_etal_1998}. The Antarctic sites such as Dome A, 
however, are particularly prone to aurora that can be extremely 
bright in the optical passbands. Broadband filters and low 
resolution spectrographs covering the auroral 
lines are sufficiently likely to be contaminated by strong 
emission lines from aurorae, i.e., the N$_2$ second positive (2P) 
and N$_2 ^+$ first negative (1N) bands dominating the $U$ and $B$ 
bands, the [\ion{O}{1}] 557.7 nm emission dominating the $V$ band, 
and the N$_2$ first positive (1P), N$_2 ^+$ Meinel (M) and O$_2$ 
atmospheric bands dominating the $R$ and $I$ bands 
\citep{Gattinger_etal_1974,Jones_etal_1975}. 
Customized filters or spectrographs with a moderately high 
resolving power can minimize the contamination from aurora and 
airglow emissions. We refer to \citet{Sims_etal_2012a} for a 
more comprehensive review of airglow and aurorae as dominant 
sources of sky brightness in Antarctica sites. 

Diffuse light from the Milky Way Galaxy could also contributes to the sky brightness. 
The Galactic Latitude $b$ of the SCP is $-27^\circ.4$, and part of the Galactic 
plane was included in the GASC FOV. The plate scale of GASC is approximately 
147$\arcsec$ per pixel, and the sub-pixel stellar contamination needs to be 
calculated and removed from the measured sky brightness data. 
Airglow, zodiacal light, and aurorae also contribute to the sky brightness. 
The intensity and frequency of occurrence of aurorae depend upon the solar 
activity. \citet{Rayleigh_1928} and \citet{Rayleigh_etal_1935} were the first 
to note a correlation between the sky brightness and the 11-year solar cycle. 
This is due to the airglow being brighter at solar maximum and fainter at solar 
minimum \citep{Krisciunas_1997, Krisciunas_etal_2007_sky}. The 10.7-cm radio flux 
of the Sun is widely used as an index of solar activity. The 2009 winter season 
occurred during solar minimum, so the sky at Dome A should have been as dark as 
other sites at solar minimum, or $B \approx$ 22.8 mag arcsec$^{-2}$ and 
$V \approx$ 21.8 mag arcsec$^{-2}$. We do not expect that the Dome A measurements 
of 2009 are significantly affected by auroral events. 

An approach to determine the sky brightness and estimate the cloud cover is given 
in the following subsections. Due to the extremely wide FOV and the fisheye optical 
design of GASC, scattered light from the edges of the optical system, as well as 
the reflection and refraction inside the optical system, is inevitable. The 
actual contribution from the Sun and the Moon cannot be well modeled 
when the sky becomes too bright. A rough model of the Sun and the Moon's 
contribution to the sky brightness will be discussed.
\vspace{-2em}
\subsection{\vspace{-1em}GASC Measurements of Sky Brightness}
The sky brightness is transformed from analog-to-digital units (ADU) into units 
of mag arcsec$^{-2}$ for each photometric band. The GASC instrumental magnitude 
is defined as: 
\begin{equation}
{m}_0 \; = \; 25 - 2.5 \; {\rm log}_{10}({\rm ADU}) \; 
\label{eqn_3}
\end{equation}
The sky brightness in units of mag arcsec$^{-2}$, which varies from band 
to band, can be defined as:
\begin{equation}
S_{\lambda} \; = a \; + \; b \times [25 \; - \; 2.5 \times {\rm log}_{10} 
({\rm ADU / pix^2})]
\label{eqn_4}
\end{equation}
Where ``pix'' is the pixel scale in unit of arcsec pixel$^{-1}$. 
The constant term in the linear calibration models is $a$, and the 
coefficient scaling the instrumental magnitude is $b$. In a certain 
sky region we wish to calibrate, we draw a box and investigate the 
statistics of the ADU values amongst all the pixels inside. We 
choose the `mode' value to best represent the sky brightness which 
is a more stable measurement as it less affected by contamination 
from the bright sources, the wide-spread PSF of stars due to the 
GASC optical system, and other unexpected events such as bright 
local aurorae. 
However, even the smallest pixel scales in GASC are 147.3 arcsec pix$^{-1}$ 
near the center of the FOV, corresponding to a box of 2.5$\arcmin
\times 2.5 \arcmin$ on the sky. 
The measured sky brightness will inevitably be contaminated by the unresolved faint sources. 

We looked at several small 
regions which lack bright sources to reduce the effect of stellar contamination. 
For instance, a box centered at RA = $2^h 24^m$, DEC = $-86^\circ 
25\arcmin$ and 25$\times$25 pixels in size ($\sim 1^\circ \times 1^\circ$) was 
inspected. The $B$-band and $R$-band magnitudes of 9550 stars in this region 
were obtained from the USNO-A2.0 catalog.  We estimated a stellar contamination of 
24.14 mag arcsec$^{-2}$ in the $B$-band. Using a mean $V$-band contamination of 
23.31 mag arcsec$^{-2}$ and a calculated median color of $V-R$ = 0.4 
mag based on the catalog from \citet{Landolt_1992}, we 
estimated the $R$-band contamination to be 22.91 mag arcsec$^{-2}$. 

\begin{figure}[!htbp]
\epsscale{0.6}
\plotone{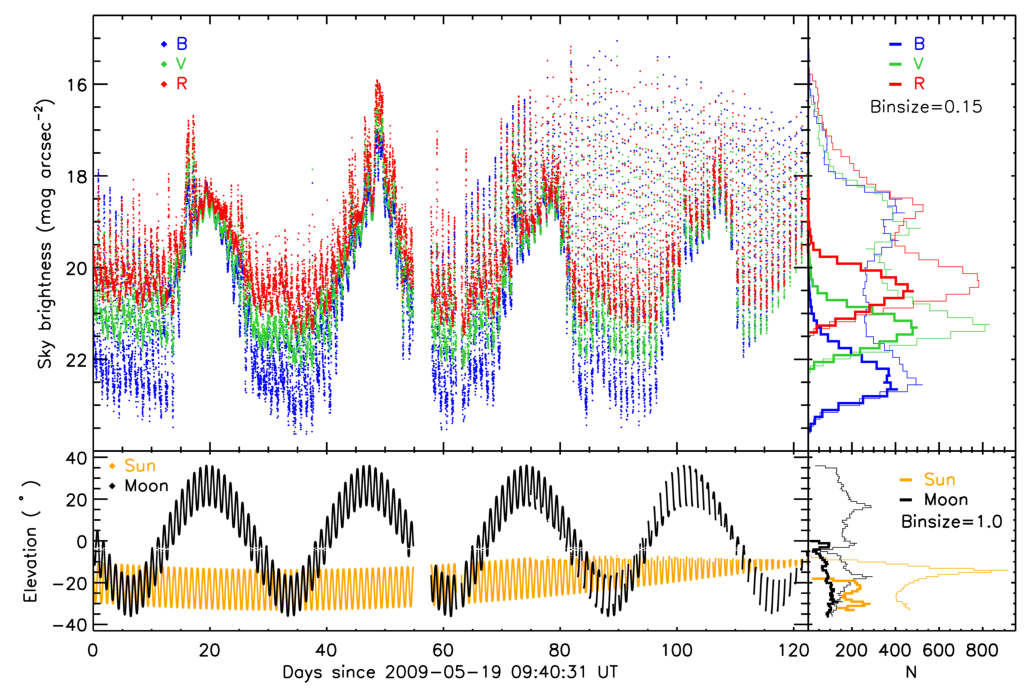}
\vspace{-1em}
\caption{Multi-band sky brightness within a 1 square degree region near the SCP, as well 
as the Sun's and Moon's elevation during the 2009 winter season. The upper and 
lower left panels present the time series while the top and bottom right-hand panels 
show the histograms. The results for the Bessell $B$-, $V$-, and $R$-bands are 
represented by blue, green, and red symbols, respectively. In the right panels, 
the histograms with solid thick lines represent the statistics for sky brightness 
during dark time, when the solar elevation angle is less than $-18^{\circ}$ and 
lunar elevation angle is less than $0^{\circ}$. Stellar contamination has already 
been removed by subtracting the contribution of a total of 9550 stars in the 
inspection area. Their magnitudes were obtained from the USNO A-2.0 
catalog. 
\label{time-series}} 
\end{figure}

\begin{figure}[!htbp]
\epsscale{0.6}
\plotone{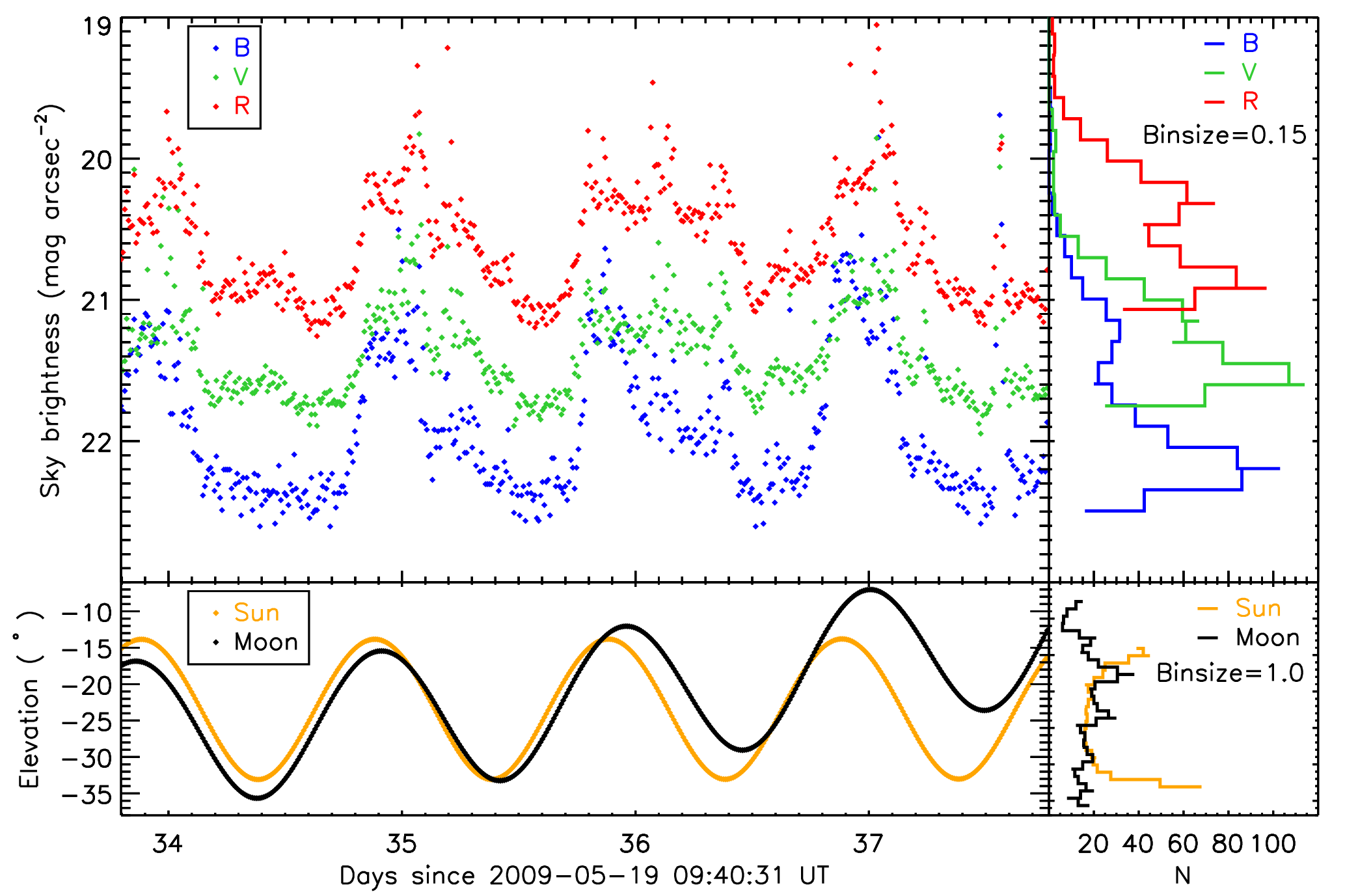}
\vspace{-1em}
\caption{A four-day subset of data shown in Fig. \ref{time-series}, from 04:25 UT 
on 2009-06-22 through 03:47 UT on 2009-06-26.  When the Moon is many degrees 
below the horizon, the daily variation of sky brightness is dominated by the 
elevation of the Sun.
\label{time-series-zoom}} 
\end{figure}

\begin{figure}[!htbp]
\epsscale{0.55}
\plotone{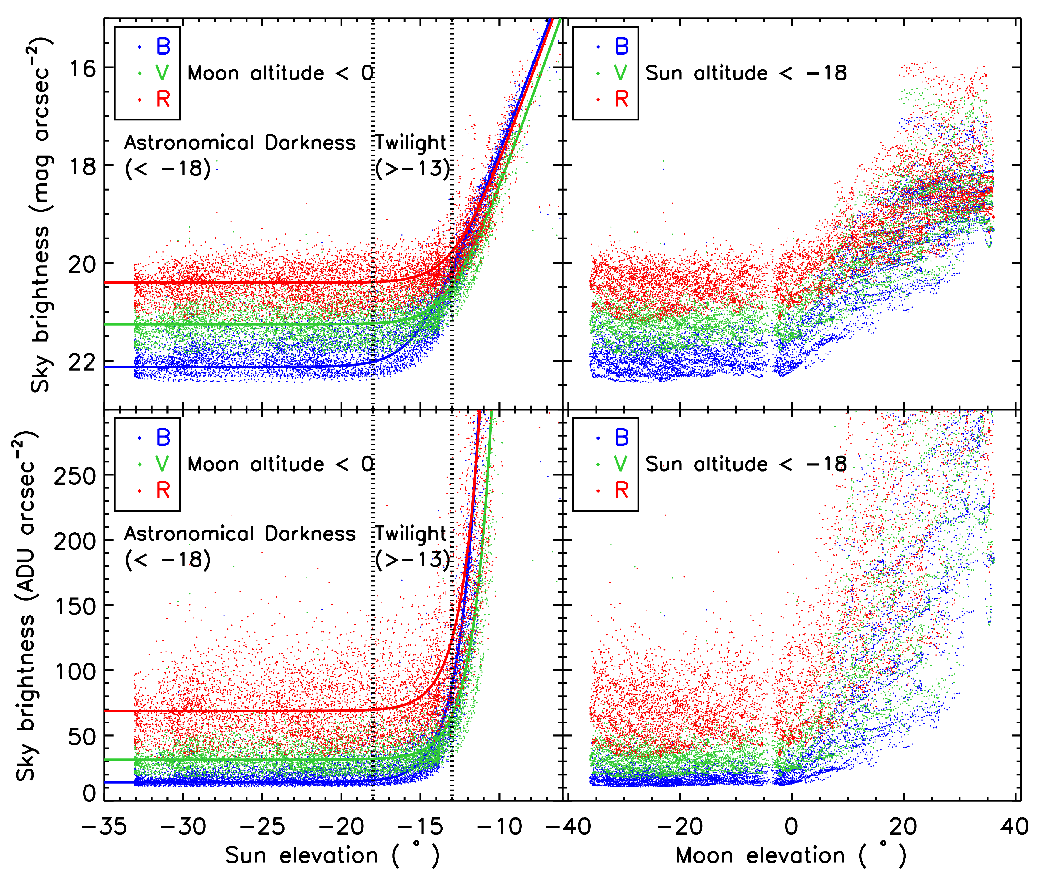}
\vspace{-1em}
\caption{Multi-band sky brightness vs. the Sun and Moon elevation. The upper panels show 
the measurements in mag arcsec$^{-2}$ while the lower panels show the data as ADU's 
per square arcsec. The left-hand panels show the relation between the sky 
brightness and the elevation angle of the Sun together with the model from 
Equation \ref{eqn_6}. Only the data with Moon elevation 
less than $0^\circ$ have been included. 
The right panels show the relation between the sky brightness and the 
elevation of the Moon. Only the data with Sun elevation less than $-18^\circ$ 
have been included.
\label{Sun-Moon-elev}} 
\end{figure}

Fig. \ref{time-series} shows the sky brightness variations during the 2009 
observing season. At such a southerly latitude as that of Dome A, the Moon is 
always fairly full when it is above the horizon from April to August, leading 
to a strong correlation between lunar elevation and sky brightness 
\citep{Zou_etal_2010}. There is a monthly variation of sky brightness which is 
strongly correlated with the lunar elevation angle. The GASC sensitivity did 
not allow data acquisition when the sky brightness was above a certain threshold. 
A dramatic enhancement in the sky brightness can be identified by looking at the 
data obtained late in the 2009 winter season. Fig. \ref{time-series-zoom} is a zoomed 
in plot for four consecutive days during the midwinter of 2009.  
In Fig. \ref{Sun-Moon-elev}, the Moon's contribution is negligible when it is 
more than $7^\circ$ below the horizon. However, a variation of the sky brightness of 
more than 1 mag arcsec$^{-2}$ can be identified which shows a strong correlation 
with the Sun's elevation angle. 
\vspace{-2em}
\subsection{\vspace{-1em}Comparison with Sky Brightness at Palomar}
Additional tests of GASC were conducted at the Palomar Observatory\footnote[4]{
Geographical coordinates of Palomar Observatory: latitude $33^\circ$ 21
$\arcmin$ 21$\arcsec$ N, longitude $116^\circ$ 51$\arcmin$ 50$\arcsec$ W.}. 
The ``ring correction'' to light curves and the fitting of calibration 
models only works based on an entire cycle of the track of the stars.  This 
allows the determination of the position within a ring where stars are least 
affected by instrumental effects.  Though it is not feasible to find the 
maximum transmission for each star cycle from tests at Palomar, we can still 
point GASC near the zenith and obtain different calibrations based on the 
instrumental magnitudes measured by GASC and the corresponding catalog 
magnitudes.

On July 5, 2013, GASC arrived at Palomar Observatory and was reassembled. 
Two tests were carried out. The first test was to compare GASC-measured sky 
brightness with Palomar NSBM measurements. We pointed the GASC at the zenith and 
set the exposure time to 50 seconds for the Bessell $B$, $V$, and $R$ 
filters\footnote[5]{For the measurements at Palomar we note a roughly 0.5 
mag arcsec$^{-2}$ variation of the sky brightness over the course of the night 
due to the band of the Milky Way passing overhead.}.  The calibration was 
carried out based on single frames of high image quality for each bandpasses. 
We used the instrumental magnitudes of the standard stars in one single 
high-quality frame per filter taken under photometric conditions.  This is 
different than the method used for data obtained at Dome A, where the brightest 
instrumental magnitudes over the course of a day were adopted as the throughput 
of the system. 

For each star in the FOV of each single exposure, the orientation of 
its maximum transmit position on the CCD chip is randomly distributed. 
In order to compare the Palomar calibration with the calibration of 
Dome A data (whose calibration models have been based on the standard 
stars' maximum transmitted flux), we performed another calibration of 
Dome A data, based on the median instrumental magnitude of each 
standard star as it tracks during one daily cycle to simulate the 
calibration that use the stars' flux at random positions like the 
Palomar test. 
By treating either the brightest or the median 
magnitude of standard stars along complete circles in Dome A data as the 
instrumental magnitude, an intrinsic offset of $(-8.670)-(-8.564)=-0.11$ 
magnitude is obtained due to the different measures of instrumental 
magnitude. The difference in the $V$-band median sky brightness on the night of 
5 July 2013 UT at Palomar Observatory, as measured by NSBM and GASC, was 
$(20.880-20.653)=0.23$ mag arcsec$^{-2}$. Thus, GASC and NSBM agree within 
-0.11 + 0.23 = 0.12 mag arcsec$^{-2}$, and the ``ring calibration'' method 
gives a reasonable calibration for the GASC data. 

Usually, inland astronomical sites are affected to some degree by artificial 
light pollution from populous cities. The sky brightness as a function of 
elevation angle obtained from the Tucson lab sites shows that there is a 
significant difference in sky brightness between the zenith and 20$^{\circ}$ 
elevation \citep{McKenna_2008}. At Cerro Tololo Inter-American Observatory 
the $V$-band sky brightness deviates from the model of \citet{Garstang_1991} 
due to light pollution at elevation angles of $\lesssim$ 10$^{\circ}$ in the 
direction of La Serena \citep{Krisciunas_etal_2010}. 
Without accounting for stellar contamination, Table \ref{percentages} presents 
the median sky brightness for different regions at Dome A, Antarctica, during 
the 2009 winter season, both for the dark time and whole season (the values 
within parentheses). Five concentric circular areas, of increasing radius and 
centered at the SCP, were inspected. 
Though the regions were centered 
at the SCP instead of the zenith, the approximate 10$^{\circ}$ offset has been 
ignored. From Table \ref{region_size}, no significant increase in brightness can 
be identified as a function of increasing angular radius.  This indicates that 
within 30$^{\circ}$ of the SCP there is dark sky that remains roughly constant 
in brightness.
\vspace{-2em}
\subsection{\vspace{-1em}Sun and Moon Model}
\citet{Liu_etal_2003} modeled the relationship between the sky brightness and the 
phase and elevation angle of the Moon. Independent to the scattering of light 
caused by reflection and refraction in the GASC optical system, the $B$-, $V$-, 
and $R$-band data should exhibit the same functional form relating to the Sun's 
and Moon's effects. 
We can write: 
\begin{equation}
{F_{Sun} \; = \; a10^{b \theta}}\; + \; c \; ,
\label{eqn_6}
\end{equation}
where F$_{Sun}$ gives the sky flux when the Moon's contribution is negligible, 
and $a$, $b$, and $c$ are constants determined for different bandpasses, 
$\theta$ is the elevation angle of the Sun. 
The multi-band sky brightness has been fitted with a nonlinear least-squares 
method using the images with good transparency and negligible contributions 
from the Moon. 

The model for the sky surface brightness due to the Moon's contribution 
involves factors such as the Earth-Moon distance and the Moon's phase. 
Following \citet{Liu_etal_2003}, the apparent magnitude of the Moon can 
be approximated by this empirical formula:
\begin{equation}
{V(R,\Phi) \; = \; 0.23 \; + \; 5\; {\rm log}_{10} R \; - \; 2.5 \; {\rm log}_{10} P(\Phi)} \; .
\label{eqn_7}
\end{equation}
where R is the Earth-Moon distance, $\Phi$ is the lunar phase angle, 
and P($\Phi$) is the function of the full Moon luminance. Following 
\citet{Zou_etal_2010}, we apply the same approach for the sky surface 
brightness contribution by the Moon. $F_{Moon}$ can be expressed as 
a form of Equation \ref{eqn_6} multiplied by the Moon phase factor 
P($\Phi$). Then,
\begin{equation}
{F_{Moon} \; = \; A P(\Phi) 10^{B \Theta} \; + \; C} \; ,
\label{eqn_8}
\end{equation}
where $\Theta$ is the elevation angle of the Moon and A, B, C are constants 
determined for each bandpass. For a more refined but slightly complicated 
sky brightness model one can consult \citet{Krisciunas_etal_1991}. The 
multi-band sky brightness has been fitted with a nonlinear least-squares 
method using images with good transparency and negligible contribution from 
the Sun. The models for the Sun's and the Moon's effect are shown in 
Table \ref{Sun_Moon_models}. 
\vspace{-2em}
\subsection{\vspace{-1em}Astronomical Twilight}
When the Sun sets, civil twilight occurs, by definition, when the Sun is 
$-12^\circ$ below horizon.  Astronomical twilight ends when the Sun reaches 
$-18^\circ$ below the horizon. If the sky brightness changes when the Sun is 
further below the horizon, it is due to changes in the airglow contribution, 
aurorae, or stellar contamination.  However, the definition of twilight 
depends not only on the photometric band pass, but also the atmospheric 
conditions at the site. Fig. \ref{Sun-Moon-elev} shows the relationship 
between the Sun and the Moon elevation on the sky brightness. 
The flux from the Moon, however, becomes significant 
only very close to the time of moonrise. Table \ref{Sun_elevation} roughly 
shows the quantitative effect of the Sun's elevation below the horizon on the 
sky brightness.

\begin{figure}[!htbp]
\epsscale{0.55}
\plotone{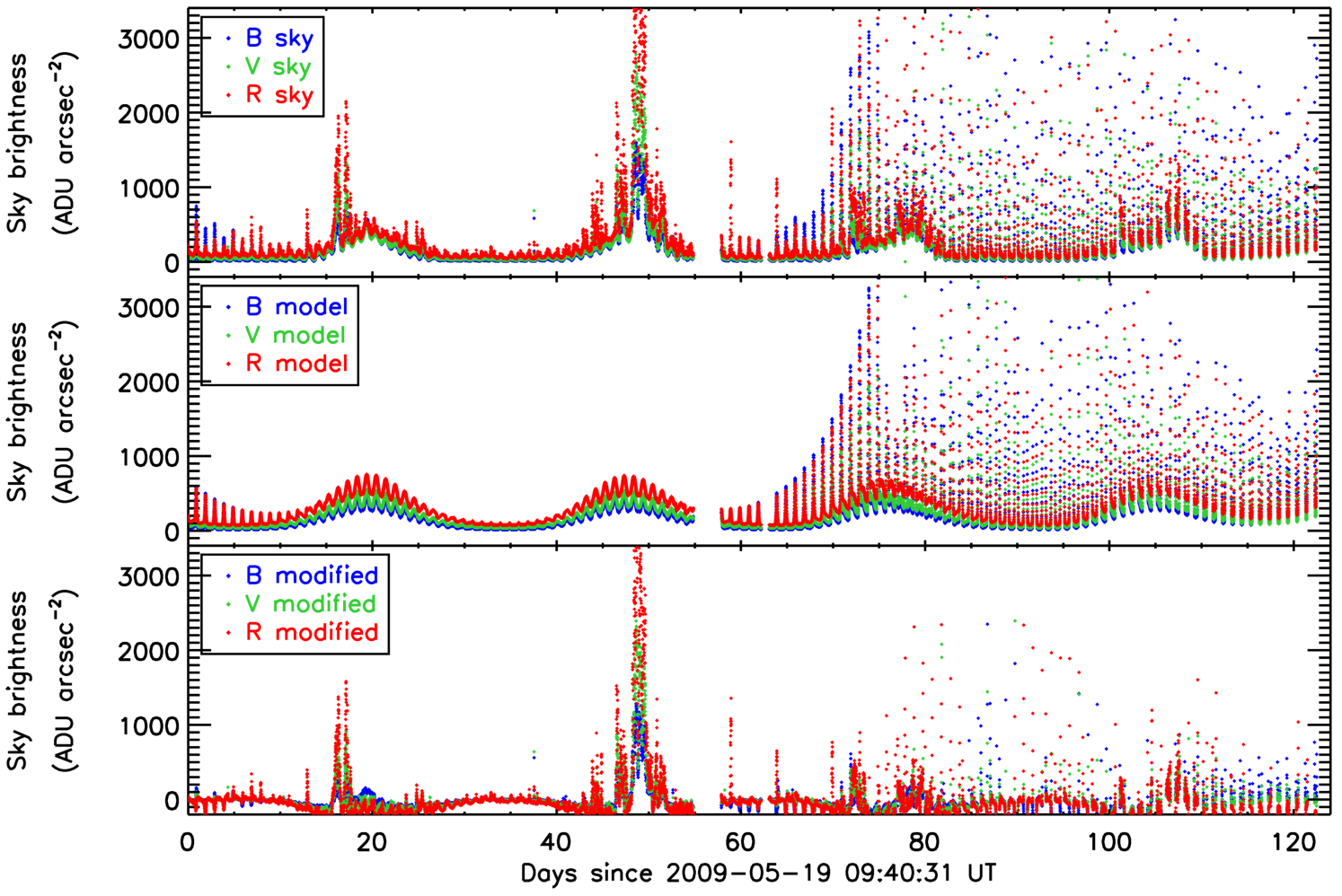}
\vspace{-1em}
\caption{Application of the sky brightness models to correct the effects of 
the Sun and the Moon. Top panel: Measured sky brightness in ADU's per square arcsec. 
Middle panel: Our Sun and Moon model in the same units. Bottom panel: Data from the 
top panel minus the Sun and Moon model shown in the middle panel. 
\label{without-with-corr}} 
\end{figure}

Fig. \ref{without-with-corr} shows the measured sky brightness in $B$, 
$V$, and $R$ (the top panel).  The middle panel shows our model of the solar 
and lunar contributions to the sky brightness. The bottom panel shows the 
observed sky brightness minus the contributions of the Sun and Moon from our 
model. The {residuals} are predictably flatter because we have subtracted 
off the contribution of the Moon when it is above the horizon. Theoretically, 
the contributions of the aurora and airglow can be estimated after properly 
removing the solar and lunar contributions to the sky background. However, 
there is still a significant fraction of scattered light that cannot be well 
modeled within the area of study $20^\circ$ in diameter, 
especially when the Moon has a higher elevation angle. 
Hence, we do not provide {any quantitative estimate} of 
aurora and airglow in our inspecting area. During the 2009 observing season 
there were few large enhancements of the sky brightness when the Sun and Moon 
had low elevation angles.  We have minimal evidence of aurorae in our data.
\vspace{-2em}
\subsection{\vspace{-1em}Extinction, Transparency Variations, and the Estimation of Cloud Cover}
The GASC FOV was centered near the SCP and extended to a zenith angle of 
40$^{\circ}$. The ``air mass'' X is the path length through the atmosphere 
at zenith angle $z$ compared to the path length at the zenith, and X = sec($z$). 
At $z$ = 40$^{\circ}$, X $\approx$ 1.3. 
At the far south latitude of Dome A any individual star within $40^{\circ}$ 
of the zenith exhibits a small range of zenith angle over the course of the night. 
GASC observed many stars at any given time over a range of 0.3 air masses. 
{Moreover,} the measurement of atmospheric extinction with GASC data is made more 
complicated by vignetting, the angular response of the interline sensor, as well 
as the different paths of light transmission through the cover window. 

Atmospheric extinction is expected to be small at Dome A. For reference, at the 
summit of Mauna Kea, Hawaii (which has a comparable elevation of 4205 m), the 
mean $B$- and $V$-band extinction values are 0.20 and 0.12 mag airmass$^{-1}$, 
respectively \citep{Krisciunas_etal_1987}.  The $R$-band extinction would be 
lower, about 0.10 mag airmass$^{-1}$.  
Let $\Delta$ be the difference of the instrumental magnitudes and the catalog 
magnitudes of stars of known brightness.  If the extinction at Dome A is 
comparable to that at Mauna Kea, over the GASC FOV we would expect $\Delta$ 
to exhibit a range vs. air mass of roughly 0.06 mag in the $B$-band, 0.04 mag 
in the $V$-band, and 0.03 mag in the $R$-band. No effect caused by the range of 
airmass has been detected with GASC data given its photometric accuracy, 
indicating a smaller atmospheric extinction coefficient at Dome A Antarctica 
compared to the Mauna Kea. 

\begin{figure}[!htbp]
\epsscale{0.75}
\plotone{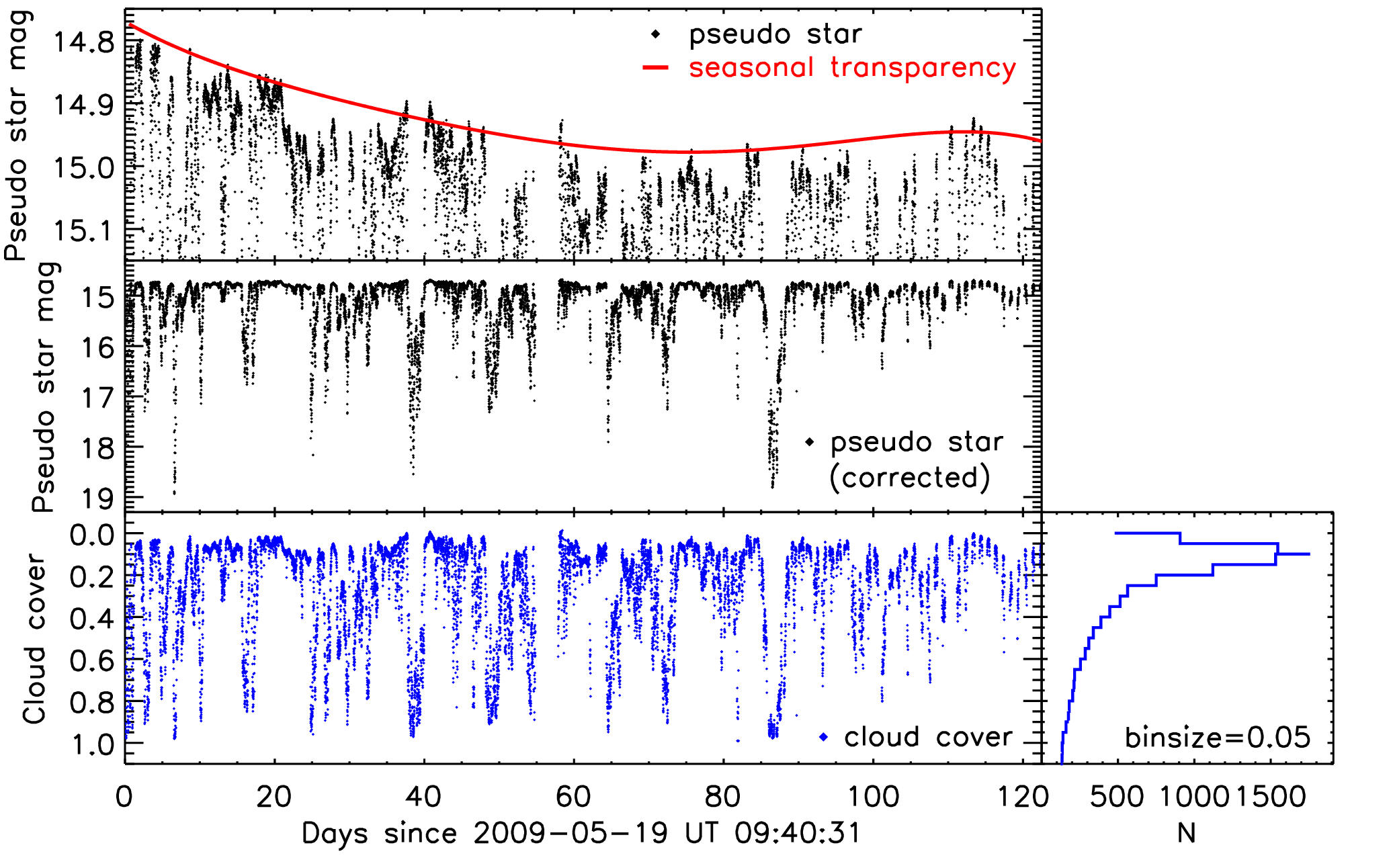}
\vspace{-1em}
\caption{The atmospheric transparency estimated from the ``pseudo-star'' after 
correction of the long-term transparency variations. The black dots in the top 
panel are intentionally plotted with a small range of brightness of the 
pseudo-star. The red curve is a polynomial fit to the upper envelope and shows 
a long-term trend in the atmospheric transparency. The middle panel shows the 
variation of the ``pseudo-star'' after removing the seasonal transparency 
variation. The lower panel shows the time-series diagram of the implied cloud 
cover, with a histogram of the cloud cover data at the right. All magnitudes 
are uncalibrated instrumental magnitudes.
\label{transparency}} 
\end{figure}

{We used the ''pseudo-star`` described in Section \ref{sec_pseudostar} as an 
indicator of the relative transparency variations to derive the likelihood 
of cloud cover at Dome A during the 2009 winter season.} 
The reduction in transparency could be due to clouds, seasonal atmospheric 
variations, or even ice formed on the entrance transmission window. Some of 
those pairs of effects can be hardly separated, as they produce the same 
effect in the change of the transparency. Therefore, our results represent 
the upper limits to the cloud cover. Fig. \ref{transparency} shows the 
transparency and the estimated cloud cover during the 2009 winter season. 
A long-term variation in transparency inferred from the brightness of the 
``pseudo-star'' is unlikely due to cloud coverage, but is more likely 
attributable to a seasonal variation of the atmosphere above Dome A. A 
fifth-order polynomial has been used to fit this long-term trend, and the 
residuals were used to calculate the upper limit of the cloud coverage. The 
estimation of the cloud coverage is also based on the ``pseudo-star'' after 
applying a correction to this long-term variation. The brightest values of 
the ``pseudo-star'' indicate very clear sky with cloud coverage estimated 
to be 0, and the reduction of the ``pseudo star'' magnitude, defined as 
$\Delta m$, was correlated with the cloud coverage as follows:
\begin{equation}
{\Delta m=-2.5 {\rm log} \frac{flux_1}{flux_2} = \; -2.5 {\rm log} (1- cloud \; cover) \; .}
\label{eqn_9}
\end{equation}
We find that the seasonal transparency degraded after June 2009, during 
which the Sun was furthest below the horizon for the year. This agrees 
with \citet{Zou_etal_2010} to some extent. However, the possibility that 
such a long-term transparency variation is due to a change in the condition 
of the instrument cannot be ruled out. Table \ref{cloud_cover} gives the 
cloud coverage percentages at Dome A from 2009-05-19 to 2009-09-18. 
A rough comparison of the cloud coverage at Mauna Kea is given in Table 
\ref{cloud_MK_ref}. This includes the cloud cover measured at the Gemini 
North Telescope and measurements with CSTAR in the $I$-band at Dome A 
during the 2008 winter season \citep{Zou_etal_2010}. CSTAR pointed at the 
SCP with a FOV of diameter $4.5^\circ$ while the GASC FOV was $85^\circ$. 
The results from 2008 and 2009 are comparable. At Dome A it is ``cloudy'' 
or worse 2\% to 3.5\% of the time, while at Mauna Kea this number is 
much higher, 30\%.  At Dome A there is less than 0.3 mag of 
extinction 62-67\% of the time, while at Mauna Kea the sky is 
photometric only 50\% of the time.

\begin{figure}[!htbp]
\epsscale{1.0}
\plotone{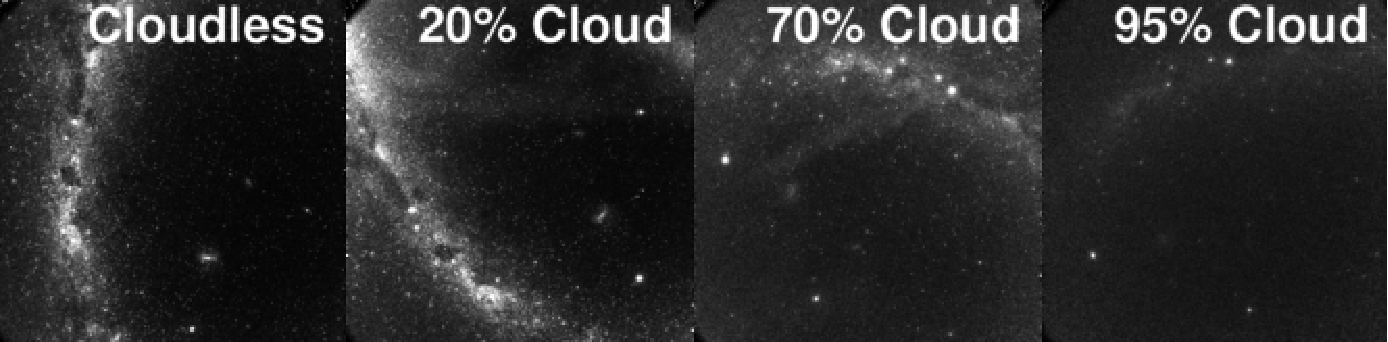}
\vspace{-2em}
\caption{Four sample images showing cloudless sky, 20 percent cloud cover, 
70 percent cloud cover and 95 percent cloud cover, from left to right, respectively.
\label{sample-clouds}} 
\end{figure}

\begin{figure}[!htbp]
\epsscale{0.6}
\plotone{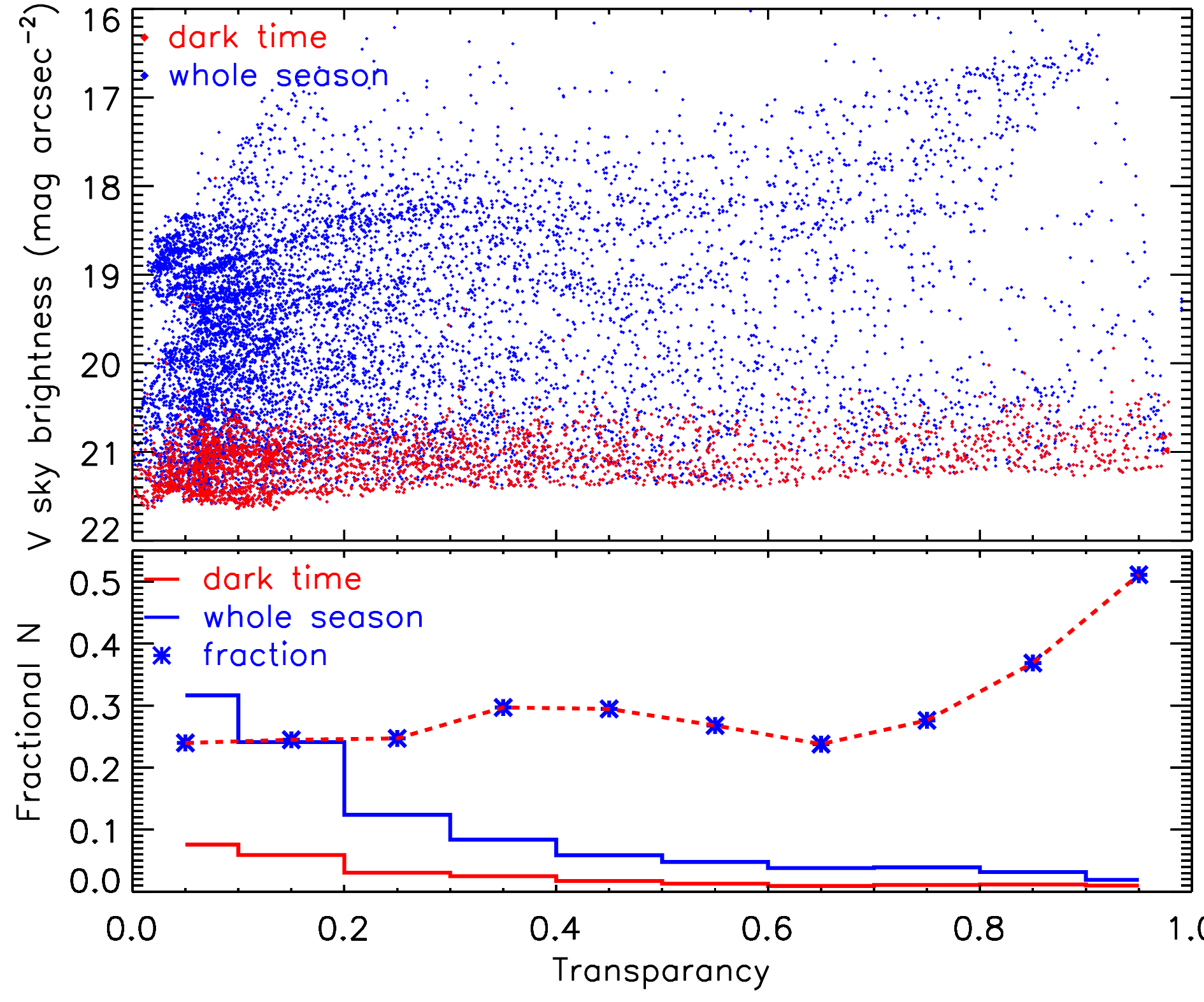}
\vspace{-1em}
\caption{The $V$-band sky brightness derived from the median ADU's within a $20^\circ$ circle
centered at the SCP vs. the transparency (upper panel). The blue and red dots 
represent the sky brightness for the entire season and during the dark time, 
respectively. The lower panel shows the normalized histograms for the $V$-band 
sky brightness. The blue asterisks with red dashed lines show the ratio of the 
bin counts of the two histograms. The bottom panel shows that the transparency 
is independent of the sky brightness in seasonal statistics
\label{SCP-Vband}} 
\end{figure}

A simple but effectively reliable way to check the cloud coverage estimated 
from the ``pseudo-star'' is to look at the original frames for certain 
fractions of cloud cover. Fig. \ref{sample-clouds} presents four sample 
images of cloud coverage of 0, 20, 70, and 95 percent obtained on 2009-06-26 
at 01:16:22, 04:10:56, 18:23:18, and 20:54:41 UT. Many images estimated 
to have high cloud cover in GASC data did not show obvious cloudy patches. 
Instead, they showed a reduction in transparency over the entire FOV. It is 
hard to determine whether those extremely low transparency events were due 
to the sky or ice formation on the entrance window. However, we can look at 
the sky brightness and the transparency estimated by the pseudo-star to see 
whether the estimation in transparency biased the sky background. Figure 
\ref{SCP-Vband} shows the transparency-sky brightness diagram. The lower 
panel shows that the transparency is independent of the sky brightness in 
seasonal statistics, indicating our estimation of the cloud coverage based 
on the pseudo-star is not biased by the different sky background.
\vspace{-2em}
\subsection{\vspace{-1em}Example Light Curves for Bright Stars}
\begin{figure}[!htbp]
\epsscale{1.0}
\plotone{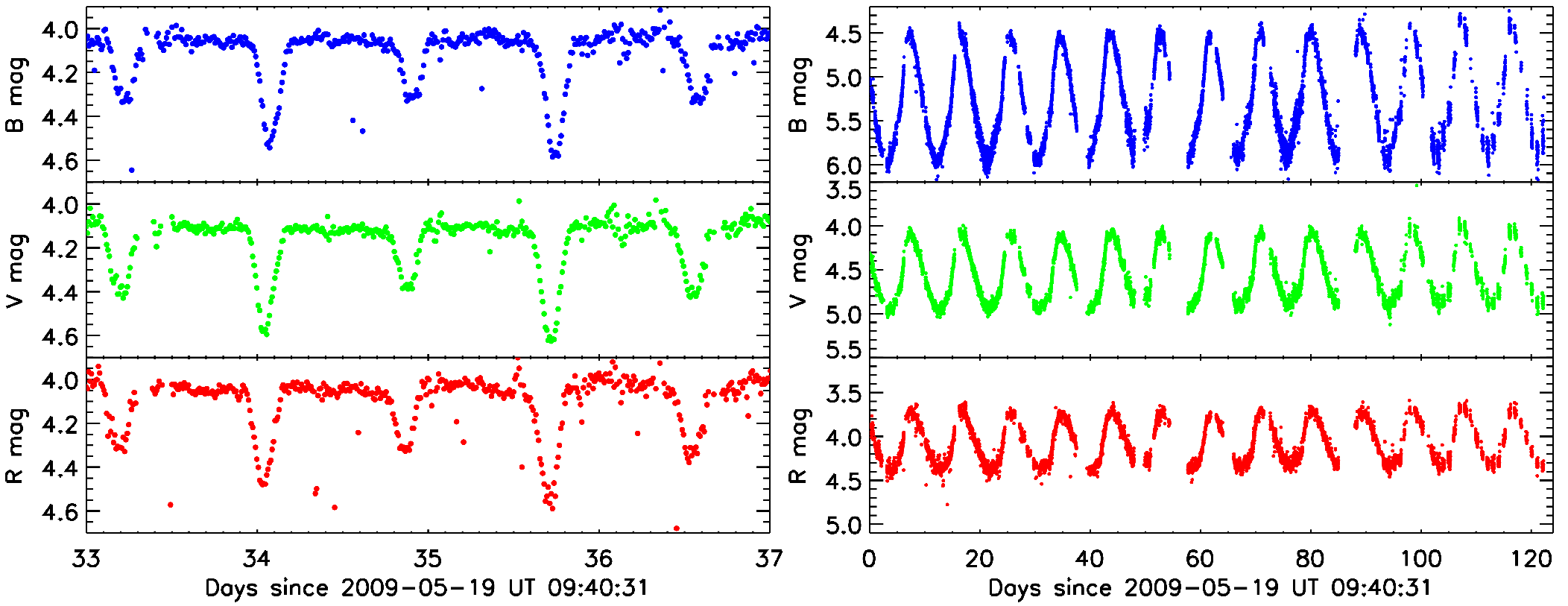}
\vspace{-1em}
\caption{The $B$, $V$, and $R$ band light curves for an eclipsing binary $\zeta$ Phoenicis
(left panel) and a W Vir type Cepheid variable $\kappa$ Pavonis (right panel). 
\label{example_lc}} 
\end{figure}

{
High-precision, high-cadence, non-consecutive time-series photometry 
serves as one of the major technical requirements for conducting 
asteroseismology. The search of exoplanets also benefits from 
high-quality photometric monitoring of stars. 
Stars within a magnitude range of $\sim$8 to $\sim$15 can be measured 
with $\sim$10 cm class and larger telescopes, however, 
consecutive monitoring of stars are even brighter, i.e., 3$-$7 magnitude, 
has not been feasible for previous Antarctic observations due to very 
short time to reach the saturation level of a detector. 

Our ``ring correction'' techinque allows us to obtain a dispersion level of 
$\sim$0.03 mag for stars around 5.5 mag in four consecutive days. 
This valuable long-term, multi-color, consecutive photometric dataset 
allows the studies on eclipsing binaries, Cepheids, and other stellar variables.
In {Figure} \ref{example_lc}, we briefly present example light curves for 
a bright eclipsing binary $\zeta$ Phoenicis and a W Vir type Cepheid 
variable $\kappa$ Pavonis {with} a short (4-day) and a long (120-day) period, 
respectively. More than 60 variables have been monitored by the GASC in 
$B$, $V$, and $R$ band. 
The multi-band photometric studies of these bright variables will be presented in another paper.} 

\vspace{-1em}
\section{\vspace{-1em}Conclusions}
In 2009 the Gattini-Dome A All-Sky Camera was deployed at Dome A in Antarctica to monitor the sky 
background, the variation of atmospheric transparency, and to perform 
photometry of bright targets in the field with an unprecedented window 
function. About 36,000 scientific images with 100 second exposure time, 
covering the Bessell $B$, $V$, and $R$ photometric bands have been used 
to quantify the $B$-, $V$-, and $R$-band sky brightness, and to estimate 
the upper limit of cloud coverage.  In a subsequent paper, we shall present 
photometry of more than 60 bright stars in our FOV that show significant 
variability based on GASC data after applying the method we 
developed to correct for the systematical error. 

The median value of sky brightness when the Sun elevation is less than 
$-18^\circ$ and the Moon is below the horizon is 22.45 mag arcsec$^{-2}$ 
for $B$-band, 21.40 mag arcsec$^{-2}$ for the $V$-band, and 20.56 mag 
arcsec$^{-2}$ for the $R$-band. If we consider a cumulative probability 
distribution, the darkest 10 percent of the time the $B$-, $V$-, and 
$R$-band sky brightness is 22.98, 21.86, and 21.68 mag arcsec$^{-2}$, 
respectively. These are comparable to the values obtained at solar minimum 
at other best astronomical sites such as Mauna Kea and the 
observatories in northern Chile. 
For future instruments operating at Dome A, 
customized filters or high spectral resolution designs 
could easily obtain better values on a more routine basis.
A test carried out with GASC at 
Palomar Observatory indicated that the GASC ``ring correction'' method 
agrees with the Palomar NSBM within 0.12 mag arcsec$^{-2}$. At Dome A the 
sky brightness is quite constant within 30$^{\circ}$ of the SCP. 

A ``pseudo-star'' was constructed based on all the stars over the FOV as an 
indicator of transparency variations. The cloud coverage during the 2009 
winter season has been estimated. We found that the seasonal transparency 
worsened in June. The transparency changed considerably in June and July 
when the Sun was at its lowest below the horizon for the year. About 
63\% of the time there was little or thin cloud coverage, using 
the same criteria for the cloud coverage adopted at the Gemini North 
Observatory at Mauna Kea, and also the cloud coverage estimation from 
CSTAR \citep{Zou_etal_2010}.

Solar and lunar models for the flux contributions to the sky background 
have been fitted, and the different flux enhancements in the sky 
background for different bandpasses have been obtained. Aurora and airglow 
are hard to quantify with GASC observations due to limited photometric 
accuracy and unexpected instrumental effects. A visual inspection of the 
sky background after removing the solar and lunar contributions indicates 
a very limited effect of  auroral events during the recent solar minimum.

\vspace{-1em}
\acknowledgments 
We thank Shri Kulkarni and Caltech Optical Observatories,
Gerard Van Belle and Chas Beichman for their financial contributions to this project. 
We are grateful to Xiaofeng Wang, Chao Wu, Ming Yang, Tianmeng Zhang, 
Yanping Zhang and Jilin Zhou for helpful discussions.  
The research is supported by the Chinese PANDA International Polar Year 
project and the Polar Research Institute of China. The project was funded 
by the following awards from the National Science Foundation Office of 
Polar Programs: ANT 0836571, ANT 0909664 and ANT 1043282. 
{The project was also supported by the Strategic Priority Research Program 
"The Emergence of Cosmological Structures" of the Chinese Academy of Sciences, 
Grant No. XDB09000000.} 
JNF acknowledges the support from the Joint Fund of Astronomy of National Natural Science Foundation of
China (NSFC) and Chinese Academy of Sciences through the grant U1231202, the NSFC grant 11673003, 
the National Basic Research Program of China (973 Program 2014CB845700 and 2013CB834900), 
{and the LAMOST FELLOWSHIP supported by Special Funding for Advanced Users, budgeted and administrated
by Center for Astronomical Mega-Science, Chinese Academy of Sciences (CAMS).} 
The operation of PLATO at Dome A is supported by the Australian Research 
Council, the Australian Antarctic Division, and the University of New South 
Wales. The authors wish to thank all the members of the 2008/2009/2010 
PRIC Dome A heroic expeditions. 

\clearpage

\begin{deluxetable}{cccc}
\tablewidth{0pc}
\tablecaption{Calibration Models\label{cal_models}}
\tablehead{ \colhead{Band} & \colhead{Dome A median mag} &
\colhead{Dome A brightest mag} & \colhead{Palomar} }
\startdata
$B$ & m$_{inst} - 9.02$ & m$_{inst} - 8.92$ & m$_{inst} - 9.52$ \\
$V$ & m$_{inst} - 8.67$ & m$_{inst} - 8.56$ & m$_{inst} - 9.00$ \\
$R$ & m$_{inst} - 9.21$ & m$_{inst} - 9.10$ &                   \\
\enddata
\end{deluxetable}

\begin{deluxetable}{clccccc}
\tablewidth{0pc}
\tabletypesize{\scriptsize}
\tablecaption{Sky Brightness for Different Percentage of Time
Value\tablenotemark{a}\label{percentages}}
\tablehead{ \colhead{Band} & \colhead{Value}\tablenotemark{b} &
\colhead{80\%} & \colhead{50\%} &  \colhead{20\%} & \colhead{10\%} &  \colhead{5\%} }
\startdata
\hline
  & mode       & 21.68 (19.17) & 21.99 (20.91) & 22.22 (21.95) & 22.31 (22.15) & 22.37 (22.26) \\
B & subtracted & 22.01 (19.20) & 22.45 (21.06) & 22.82 (22.40) & 22.98 (22.70) & 23.10 (22.90) \\
  & corrected  & 22.13 (19.32) & 22.57 (21.18) & 22.94 (22.52) & 23.10 (22.83) & 23.22 (23.02) \\
\hline
  & mode       & 20.93 (19.05) & 21.22 (20.61) & 21.48 (21.24) & 21.59 (21.43) & 21.67 (21.56) \\
V & subtracted & 21.07 (19.08) & 21.40 (20.70) & 21.72 (21.42) & 21.86 (21.65) & 21.96 (21.81) \\
  & corrected  & 21.19 (19.20) & 21.52 (20.83) & 21.84 (21.54) & 21.98 (21.77) & 22.08 (21.93) \\
\hline
  & mode       & 20.13 (18.69) & 20.44 (19.91) & 20.75 (20.49) & 20.90 (20.70) & 20.99 (20.85) \\
R & subtracted & 20.21 (18.71) & 20.56 (19.98) & 20.91 (20.61) & 21.68 (20.85) & 21.20 (21.03) \\
  & corrected  & 20.34 (18.54) & 20.68 (20.10) & 21.03 (20.73) & 21.02 (20.97) & 21.15 (21.32) \\
\enddata
\tablenotetext{a}{Values without parentheses are for dark time.  Values in parentheses
are for the whole season.}
\tablenotetext{b}{mode: the `mode' value amongst all the pixels inside the inspected region; 
subtracted: the `mode' value subtracted for the stellar contaminations; `corrected': the 
`subtracted' {values} further corrected for the offset between the GASC and Palomar NSBM. }
\end{deluxetable}

\begin{deluxetable}{cccc}
\tablewidth{0pc}
\tablecaption{Mode of Sky Brightness for Regions of Different Angular 
Size\tablenotemark{a}\label{region_size}}
\tablehead{ \colhead{Diameter (deg)} & \colhead{$B$} &
\colhead{$V$} & \colhead{$R$} }
\startdata
4.6 & 21.92 (20.41) & 21.16 (20.25) & 20.40 (19.65) \\
20  & 21.90 (20.40) & 21.16 (20.27) & 20.39 (19.66) \\
40  & 21.90 (20.41) & 21.17 (20.30) & 20.40 (19.69) \\
60  & 21.96 (20.46) & 21.24 (20.37) & 20.47 (19.77) \\
\enddata
\tablenotetext{a}{Values without parentheses are for dark time.  
Values in parentheses are for the whole season.}
\end{deluxetable}

\begin{deluxetable}{ccc}
\tablewidth{0pc}
\tabletypesize{\scriptsize}
\tablecaption{Sun and Moon Models for Sky Brightness\label{Sun_Moon_models}}
\tablehead{ \colhead{Band} & \colhead{Sun Model} & \colhead{Moon Model} }
\startdata
B  & $\mathrm{F_{Sun}=2.076\times 10^{6} \times 10^{0.342 \theta} +16.283}$ & $\mathrm{F_{Moon}=118.098\times P(\Phi)10^{0.017 \Theta} -18.544}$ \\
V  & $\mathrm{F_{Sun}=1.596\times 10^{6} \times 10^{0.360 \theta} +35.463}$ & $\mathrm{F_{Moon}=151.629\times P(\Phi)10^{0.015 \Theta} -26.084}$ \\
R  & $\mathrm{F_{Sun}=2.158\times 10^{6} \times 10^{0.353 \theta} +75.622}$ & $\mathrm{F_{Moon}=232.785\times P(\Phi)10^{0.013 \Theta} -31.993}$ \\
\enddata
\end{deluxetable}

\begin{deluxetable}{cccc}
\tablewidth{0pc}
\tablecaption{Sun Elevation Angles Corresponding to Increased Sky Brightness\label{Sun_elevation}}
\tablehead{ \colhead{Flux Increase} & \colhead{$B$} & \colhead{$V$} & \colhead{$R$} }
\startdata
$20\%$  & -17.2$^\circ$ & -15.0$^\circ$ & -14.7$^\circ$ \\
$50\%$  & -16.0$^\circ$ & -13.9$^\circ$ & -13.6$^\circ$ \\
$100\%$ & -15.1$^\circ$ & -13.1$^\circ$ & -12.7$^\circ$ \\
$200\%$ & -14.2$^\circ$ & -12.2$^\circ$ & -11.9$^\circ$ \\
\enddata
\end{deluxetable}

\begin{deluxetable}{cccccc}
\tablewidth{0pc}
\tabletypesize{\scriptsize}
\tablecaption{Cloud Cover at Dome A\label{cloud_cover}}
\tablehead{ \colhead{Flux} & \colhead{Extinction (mag)} & \colhead{GASC2009} 
& \colhead{GASC2009$^a$} & \colhead{Cstar2008} & \colhead{Description} }
\startdata
$<50\%$     & $>0.75$       & $17.2\%$ & $19.9\%$ & $9\%$  & Thick \\
$50\%-75\%$ & $0.31 - 0.75$ & $19.4\%$ & $27.2\%$ & $17\%$ & Intermediate \\
$75\%-90\%$ & $0.11 - 0.31$ & $29.1\%$ & $42.1\%$ & $23\%$ & Thin \\
$>90\%$     & $<0.11$       & $34.3\%$ & $10.8\%$ & $51\%$ & Little or none \\
\enddata
\tablenotetext{a}{Values obtained without correcting for the long-term 
transparency variation.}
\end{deluxetable}

\begin{deluxetable}{cccccc}
\tablewidth{0pc}
\tabletypesize{\scriptsize}
\tablecaption{Cloud Cover Compared to Mauna Kea\label{cloud_MK_ref}}
\tablehead{ \colhead{Description} & \colhead{Extinction (mag)} & \colhead{Mauna Kea} 
& \colhead{DomeA (GASC2009)} & \colhead{Dome A (GASC2009)$^a$} & \colhead{Dome A (Cstar2008)} }
\startdata
Any other usable & $>3$    & $10\%$ & $1.0\%$   & $1.1\%$  & $0\%$ \\
Cloudy           & $2-3$   & $20\%$ & $2.5\%$   & $2.8\%$  & $2\%$ \\
Patchy cloud     & $0.3-2$ & $20\%$ & $34.2\%$  & $45.1\%$ & $31\%$ \\
Photometric      & $<0.3$  & $50\%$ & $62.4\%$  & $51.0\%$ & $67\%$ \\
\enddata
\tablenotetext{a}{Values obtained without correcting for the long-term 
transparency variation.}
\end{deluxetable}

\clearpage
\parindent = 0 mm


\end{document}